\documentclass[conference]{IEEEtran}
\usepackage{xspace,multirow,amssymb,tabularx} 
\usepackage{color,cite,lastpage,amsmath,graphicx,caption,subcaption}
\usepackage[table,xcdraw]{xcolor} 
\usepackage[font={small}]{caption}
\usepackage[hyphens]{url}

\def\shownotes{1}   	
\def\showedits{0} 	

\ifnum\shownotes=1 \newcommand{\authnote}[2]{{ $\ll$\textsf{\footnotesize #1
notes: #2}$\gg$}} \else \newcommand{\authnote}[2]{} \fi

\ifnum\showedits=1 \newcommand{\edit}[1]{\textcolor{red}{#1}}
\else \newcommand{\edit}[1]{\textcolor{black}{#1}} \fi

\providecommand{\systemname}{Astoria\xspace}
\providecommand{\vs}{vs. } \providecommand{\ie}{\emph{i.e.,} }
\providecommand{\eg}{\emph{e.g.,} } 
 
\providecommand{\etal}{\emph{et al. }}   
\providecommand{\etc}{\emph{etc.}}      

\providecommand{\myparab}[1]{\smallskip\noindent\textbf{#1} }

 \newcounter{GraphCounter} 

\newcommand{\squishenum}{ \begin{enumerate}{} { \setlength{\itemsep}{0pt}
\setlength{\parsep}{0pt} \setlength{\topsep}{3pt}
\setlength{\partopsep}{0pt} \setlength{\leftmargin}{1.5em}
\setlength{\labelwidth}{1em} \setlength{\labelsep}{0.5em} } }
\newcommand{\squishlist}{ \begin{list}{$\bullet$} { \setlength{\itemsep}{0pt}
\setlength{\parsep}{3pt} \setlength{\topsep}{3pt}
\setlength{\partopsep}{0pt} \setlength{\leftmargin}{1.5em}
\setlength{\labelwidth}{1em} \setlength{\labelsep}{0.5em} } }
\newcommand{\squishlisttwo}{ \begin{list}{$\bullet$} { \setlength{\itemsep}{0pt}
\setlength{\parsep}{0pt} \setlength{\topsep}{0pt}
\setlength{\partopsep}{0pt} \setlength{\leftmargin}{2em}
\setlength{\labelwidth}{1.5em} \setlength{\labelsep}{0.5em} } }
\newcommand{\squishend}{ \end{list}  }
\newcommand{\squishenumend}{ \end{enumerate}	}
\makeatletter
\def\@copyrightspace{\relax}
\makeatother
\title{Measuring and mitigating AS-level adversaries against Tor}
\begin{document}

\author{\IEEEauthorblockN{Rishab Nithyanand\IEEEauthorrefmark{1},
Oleksii Starov\IEEEauthorrefmark{1},
Adva Zair\IEEEauthorrefmark{2}, 
Phillipa Gill\IEEEauthorrefmark{1} and
Michael Schapira\IEEEauthorrefmark{2}}
\IEEEauthorblockA{\IEEEauthorrefmark{1}Stony Brook University\\
 Email: \{rnithyanand, ostarov, phillipa\}@cs.stonybrook.edu}
\IEEEauthorblockA{\IEEEauthorrefmark{2}Hebrew University of Jerusalem\\
Email: \{adva.zair@mail, schapiram@cs\}.huji.ac.il}}


\maketitle

\section*{Abstract}

The popularity of Tor as an anonymity system has made it a popular target for a
variety of attacks. We focus on traffic correlation attacks, which are no
longer solely in the realm of academic research with recent revelations about
the NSA and GCHQ actively working to implement them in practice. 

Our first contribution is an empirical study that allows us to gain a high
fidelity snapshot of  the threat of traffic correlation attacks in the wild. We
find that up to 40\% of all circuits created by Tor are vulnerable to attacks by
traffic correlation from Autonomous System (AS)-level adversaries, 42\% from 
colluding AS-level adversaries, and 85\% from state-level adversaries. In 
addition, we find that in some regions (notably, China and Iran) there exist 
many cases where over 95\% of all possible circuits are vulnerable to 
correlation attacks, emphasizing the need for AS-aware relay-selection. 

To mitigate the threat of such attacks, we build \systemname--an AS-aware Tor
client. \systemname leverages recent developments in network measurement to 
perform path-prediction and intelligent relay selection. \systemname reduces 
the number of vulnerable circuits to 2\% against AS-level adversaries, 
under 5\% against colluding AS-level adversaries, and 25\% against state-level 
adversaries. In addition, \systemname load balances across the Tor network so 
as to not overload any set of relays.

\section{Introduction}\label{sec:intro}

Tor is a popular anonymity system for users who wish to access the Internet
anonymously or circumvent censorship~\cite{Dingledine-USENIX04}. The increasing
popularity of Tor has recently made it a high-value target for blocking and
denial of service~\cite{Danner-TISSEC12, Winter-FOCI12, Hopper-WPES09} and
traffic correlation attacks to deanonymize users~\cite{Murdoch-SP05,
Shmatikov-ESORICS06,Murdoch-PETS07,Houmansadr-NDSS11,Johnson-CCS13}.  Traffic
correlation attacks, which correlate traffic entering the Tor network with
traffic exiting it, are no longer solely in the realm of academic research with
recent revelations about the NSA and GCHQ actively working to implement them in
practice, in collusion with Internet Service Providers (ISPs)\cite{Schneier-NSA1,
Guardian-NSA2, Guardian-NSA3}.  

Traffic correlation attacks have been shown to be feasible and practical for
network-level attackers. Specifically, a traffic correlation attack may be
implemented by any autonomous system (AS) that lies on both the path from the
Tor client to the entry relay and on the path from the exit relay to the
destination. \edit{Previous studies have demonstrated the potential for this type of
attack \cite{Johnson-CCS13,Edman-CCS09,Feamster-WPES04}. Proposed defenses
include relay selection strategies to avoid ASes that are in a position to 
launch them \cite{Akhoondi-SP12}.} However, recent work \cite{Wacek-NDSS13} 
has shown that these strategies perform poorly in practice. 

The threat of network-level adversaries has been exacerbated by a recent study
which highlights that the set of ASes that are in a position to perform traffic
correlation analysis is potentially much larger due to asymmetric routing,
routing instabilities, and intentional manipulations of the Internet's routing
system \cite{Vanbever-HotNets14,Sun-Arxiv15}. 
These attacks significantly raise the bar for relay-selection systems.
Specifically, they require the relay-selection system be able to accurately
measure or predict network paths in both the forward and reverse direction.
Measuring the reverse path between two Internet hosts is non-trivial, especially
when the client does not have control over the destination, as is commonly the
case for popular Web services. While solutions for measuring reverse paths have
been proposed~\cite{reverse-traceroute}, they are still not widely deployed or
available. 

In this paper, we make contributions in two dimensions. First, we quantify the 
threat posed by these new attacks. Second, we develop a relay selection method 
to minimize their impact. 

\myparab{Measuring the threat faced by Tor. }We leverage up-to-date maps of
the  Internet's topology \cite{Giotsas-IMC14} combined with algorithmic
simulations \cite{Gill-CCR12} to predict which ASes are in a position to perform
traffic correlation analysis on forward or reverse paths. We validate this
technique and show that it provides a reasonable \edit{\emph{estimate}} on the 
threat faced from AS-level attackers. We then augment our
analysis with  techniques to identify ASes owned by a single organization
(sibling ASes) in order to gain a clearer picture of which ASes are likely to
collude with each other. This provides a more complete picture of
network-level threats than previous work. In addition, we consider the threat
from state-level attackers that have insight into traffic transiting through
all regional ASes. Through these techniques and our experiments, we make the 
following key observations:

\begin{itemize}

\item Up to 40\% of circuits constructed by the current Tor client are
vulnerable to network-level attackers.

\item \edit{Up to 37\% of all sites in our study, when loaded from Brazil, China,
Germany, Spain, France, England, Iran, Italy, Russia, and the United States had
main page requests that were reached via a \emph{vulnerable} path (i.e., a path that
contained network-level entities in a position to launch traffic correlation
attacks), when loaded by the vanilla Tor client.}

\item Connections from China were found to be most vulnerable to network-level
attackers with up to 86\% of all Tor circuits and 56\% of all main page
requests to sites in the study being vulnerable to colluding network-level 
attackers.

\item \edit{For up to 8\% of the requests generated from China and Iran, over 95\% of 
\emph{all possible} Tor constructed circuits were vulnerable to correlation 
attacks by network-level attackers. }

\item \edit{Reducing the number of entry guards can result in an increase in
vulnerability of Tor circuits. In particular, we found that using a
single guard significantly increases the threat from traffic correlation
attacks, while the difference between using two and three guards is marginal.}

\item State-level attackers are in a position to launch correlation attacks on
up to 85\% of all Tor constructed circuits.
\end{itemize}

\myparab{Mitigating the threat of AS-level adversaries. }
We propose, construct, and evaluate \systemname -- an AS-aware Tor client that
includes security and relay bandwidth considerations  when creating Tor
circuits. \systemname is the first AS-aware Tor client to consider the recently
proposed asymmetric correlation attacks~\cite{Vanbever-HotNets14, Sun-Arxiv15}.
When there are safe alternatives, \systemname actively avoids using circuits on
which asymmetric correlation attacks might be launched. It also leverages
methods for identifying sibling ASes \cite{Anwar-TR15} when determining whether or
not a given circuit is safe. In the absence of a safe path, \systemname uses a
linear program to minimize the threat posed by any adversary. Finally,
\systemname considers the bandwidth capabilities of relays while making AS-aware
relay selection decisions. \edit{When there are multiple safe relay selections, 
\systemname aims to be a good network citizen and distributes load across Tor 
relays in the same manner as the vanilla Tor client.}
Therefore, in spite of selecting safer relays, \systemname will not overload any
single set of relays.

\myparab{Paper outline. }In Section \ref{sec:background} we briefly overview how
the current Tor client performs relay selection and circuit construction,
describe the current state of research in relay selection for Tor, and introduce
our adversary model. In Section \ref{sec:measurement} we describe the components
of our measurement toolkit used for detecting network-level attackers on Tor
circuits. We then present some interesting results regarding the vulnerability 
of Tor constructed circuits and the general potential for attack by
single AS-, sibling AS-, and state-level attackers. In Section \ref{sec:system}, 
we present the details of our AS-aware client -- \systemname. A performance 
and security evaluation of \systemname is performed in Section 
\ref{sec:evaluation}. In Section \ref{sec:discussion}, we discuss the
\edit{known} shortcomings of \systemname and motivate directions for future 
research on AS-aware clients. We make our conclusions in Section 
\ref{sec:conclusions}.

\section{Background and Motivation}\label{sec:background}

We now provide background on Tor relay selection, related work in this area, 
and our adversary model.

\subsection{Tor relay selection}\label{subsec:tor-relay-selection}

The Tor anonymity network consists of approximately 6,000 relays (Tor routers). Most 
requests made through a Tor client are sent to their destination via a three-hop 
path known as a circuit. Each circuit consists of an entry, middle, and exit 
relay. The entry-relay communicates directly with the client and the exit-relay 
communicates with the destination. The fundamental idea is that no single relay 
in the circuit learns the source and destination. 

In its early days, Tor selected relays for each circuit hop uniformly
at random from the set of available relays. This was changed in order to
improve performance (by preferring to route through higher bandwidth relays
\cite{move-to-higher-bw}) and security \cite{Borisov-CCS07}. In today's Tor
network, based on certain performance characteristics such as reliability, 
bandwidth served, and up-time, relays may earn certain flags that make them 
a preferential choice for various roles during circuit construction.

One such flag is the guard flag. New relays joining the Tor network are
monitored for stability and performance via remote measurements for a period of
up to eight days \cite{tor-blog-lifecycle}. At this point, relays that have
demonstrated stability and reliability are assigned a guard flag. Relays with a 
guard flag earn the ability to serve as the entry-relay to the Tor network. By 
default the Tor client selects three guards to be used as entry-relays for all 
circuits for a prolonged period of time. The main ideas behind the selection of 
a fixed set of entry-relays are (1) to reduce the possibility that a client will 
select an entry- and exit-relay operated by the same entity (after prolonged
use), (2) prevent attacker-owned entry-relays from denying service to clients 
that are not also using an exit-relay owned by the attacker, and (3) increase 
the cost to an attacker that wishes to be chosen as an entry-relay, by requiring 
them to earn the guard flag \cite{tor-blog-lifecycle}.

In addition to picking relays that are more stable and reliable, for other
locations on a circuit, the Tor client also requires that (1) no two routers on 
a circuit share the same $/16$ subnet and (2) no routers in the same 
family (as advertised by the router) may be chosen on the same circuit. 
\cite{move-to-higher-bw}.

\subsection{Related work}\label{subsec:related}

The threat of correlation attacks by AS-level adversaries on the Tor network was 
first identified and empirically evaluated by Feamster and Dingledine 
\cite{Feamster-WPES04} in 2004, when the Tor network had only 33 relays and 
significantly different relay selection algorithms. The study revealed that 
 10-30\% of all circuits constructed by Tor had a common AS that 
could observe both ends of the circuit. 
Shortly after, by constructing efficient traffic correlation attacks while 
considering network-level adversaries, Murdoch and Danezis \cite{Murdoch-SP05} 
and Murdoch and Zieli\'{n}ski \cite{Murdoch-PETS07} demonstrated that the threat 
from AS-level attackers was one of practical concern.
In 2009, Edman and Syverson \cite{Edman-CCS09} found that the threat of AS-level
adversaries had not reduced since \cite{Feamster-WPES04}, in spite of revised 
relay selection strategies and substantially larger number of relays in the 
network.

In addition, Edman and Syverson \cite{Edman-CCS09} were the first to consider 
threats from network-level attackers due to the asymmetric nature of Internet 
routing. Using the 2009 topology of the Internet, AS paths inferred by Qiu's
algorithm \cite{qiu-bgp}, and AS relationships inferred by Gao's algorithm
\cite{gao-as-relationships} they found that in their 
experiments up to 39\% of all Tor circuits were vulnerable to network-level 
adversaries that performed attacks on forward- and reverse-paths. Most recently, 
Vanbever \etal\cite{Vanbever-HotNets14} and Sun \etal \cite{Sun-Arxiv15}, 
presented RAPTOR, an AS-level attack integrating BGP interception with the
first correlation attack that takes advantage of the asymmetric nature of 
Internet routing, to exactly de-anonymize Tor users with up to 90\% accuracy 
in just 300 seconds. Similarly, Johnson \etal\cite{Johnson-CCS13} performed 
an empirical evaluation of the effect of network-level adversary bandwidth 
investment strategies, Tor client location, and Tor client use (\eg for IRC, 
browsing, BitTorrent, \etc). They found that a network-level adversary could 
effectively de-anonymize most Tor users within six months with very low 
bandwidth costs. These works emphasize the need for Tor relay selection 
strategies to consider ASes that lie both, on the forward- and reverse-paths 
between the (client, entry) and (exit, destination).

Perhaps most closely related to our work, in terms of end-goals and evaluation
methodology, Akhoondi \etal\cite{Akhoondi-SP12}, constructed LASTor, a Tor 
client which explicitly considered AS-level attackers and relay locations while 
constructing Tor circuits. While LASTor appeared to successfully reduce path 
latencies and the probability of common ASes at either end of the Tor circuits, 
it neglected the capacity of relays selected by the system. Relay capacity is 
an important variable to consider to ensure that custom relay selection schemes 
do not overload a small set of relays, therefore reducing the performance of the 
entire network. Their evaluation, based on only HTTP HEAD requests (as opposed 
to complete webpage loads), did not stress the system sufficiently to reveal the 
issues associated with capacity-agnostic relay selection. Further, LASTor does 
not consider an adversary that may (1) collude with other ASes or operate at
the state-level, and/or (2) only need to be on one of the asymmetric path 
segments between source and entry-relay; and exit-relay and destination 
(\eg RAPTOR).

\subsection{Adversary model}\label{subsec:adversary}

\begin{figure}[t]
\centering
\includegraphics[width=0.45\textwidth]{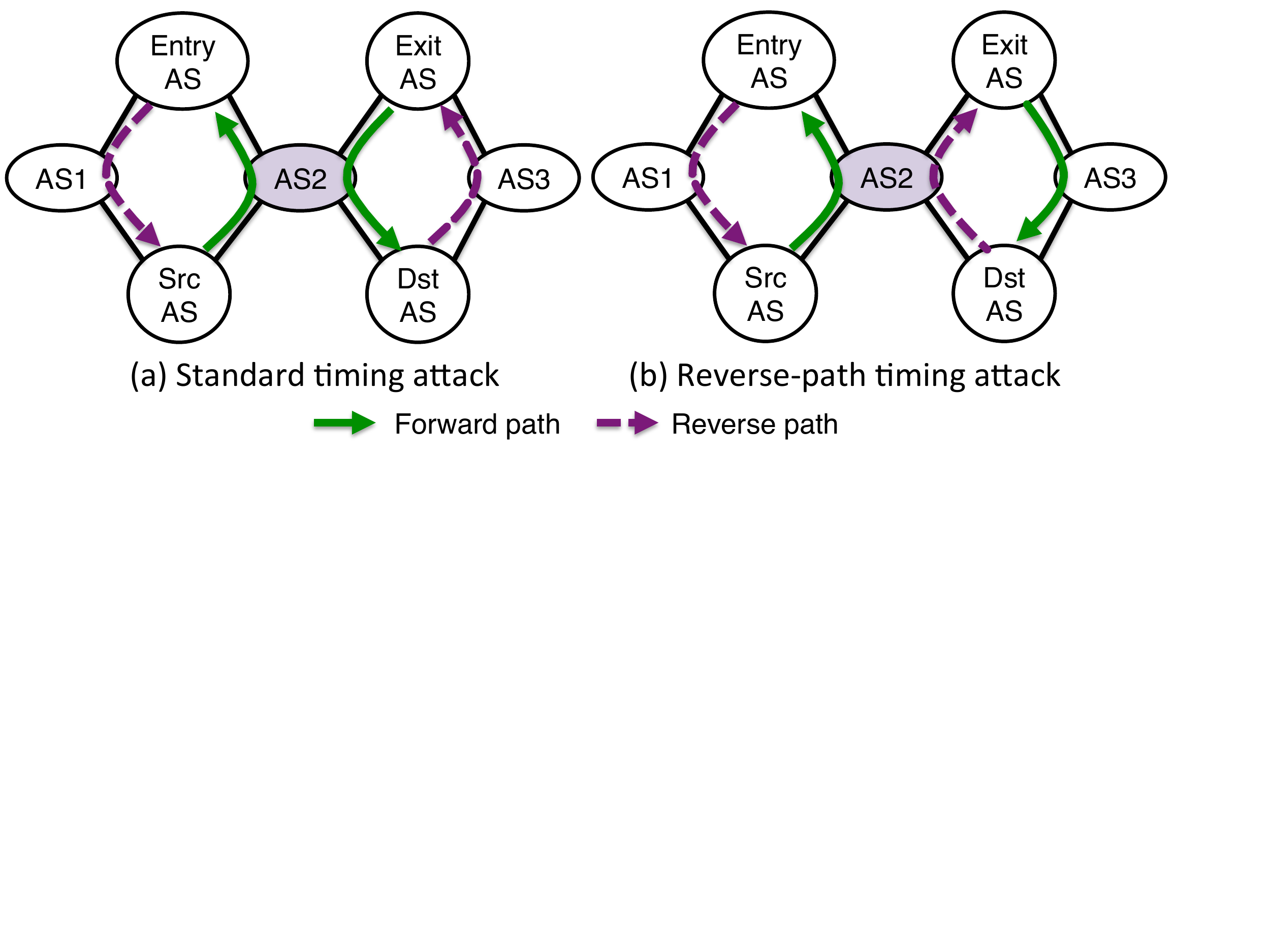}
\caption{Standard and reverse-path traffic correlation attacks. In the standard 
traffic correlation attack, AS2 must observe the direction of the connection 
that data is flowing on (forward path). In the reverse-path traffic correlation 
attack AS2 can infer the data flow using ACK numbers on the reverse path.}
\label{fig:advmodel}
\end{figure}

In the standard view of traffic correlation attacks, an AS needs to lie on the 
forward path\footnote{Here we use `forward path' to refer to the direction of 
data flow in the TCP connection} between the source and destination (i.e., on 
the solid green colored path segments in Figure~\ref{fig:advmodel} (a)). With 
this point the adversary (AS 2) can view the packet sizes and timings as 
transmitted from the source to destination, going-into and coming-out-of the 
Tor network and directly perform a traffic correlation attack. 

However, recent work by Vanbever \etal\cite{Vanbever-HotNets14} and Sun \etal
\cite{Sun-Arxiv15} highlights the fact that an adversary on the reverse path 
may also learn packet size and timing information via the TCP Acknowledgement 
(ACK) field. Figure \ref{fig:advmodel}(b) illustrates this case.  AS 2 
can directly observe packet timings between the source and entry-relay AS 
(Entry AS), but can only observe ACKs from the destination back to the 
exit-relay AS (Exit AS). 

In this view, an adversary has the potential to launch a traffic correlation 
attack on a Tor circuit as long as the following criteria are satisfied:

Let $p_{src \rightarrow entry}=\{ AS_1, AS_2, \dots, AS_n\}$ be the set of ASes 
on the path from the source (Tor client) to the selected entry-relay (this set 
includes the entry-relay AS), 
$p_{entry \rightarrow src} = \{ AS'_1, AS'_2, \dots, AS'_m\}$ 
be the set of ASes on the path from the entry-relay back to the source, and
$p_{entry \leftrightarrow src} = p_{entry \rightarrow src} \cup 
p_{src \rightarrow entry}$. We similarly define paths to and from the exit-relay and destination (\eg a 
popular content provider, or other Web service) as 
$p_{exit \rightarrow dst}$, $p_{dst \rightarrow exit}$, and 
$p_{exit \leftrightarrow dst}$. 

We say that a Tor circuit is \edit{vulnerable to a traffic correlation attack 
if there exists an AS $A_i$ such that:}
\begin{equation}\label{eq:attacked}
A_i \in  \{p_{src \leftrightarrow entry}  \cap  p_{exit \leftrightarrow dst} \}
\end{equation}

Similar to prior work on relay selection, we assume that our adversary is an 
autonomous system (AS), or an entity working with the cooperation of ASes (\eg 
governments). However, while all previous work only considers the standard view 
of network attacks, we also consider attackers that may  lie on the 
reverse-path, as described above. In addition, we also include the possibility 
that some sets of ASes may collude with each other to de-anonymize Tor users. 
Specifically, we consider that an AS may collude with sibling ASes 
\cite{Anwar-TR15} (i.e., other ASes owned by the same organization) and ASes
that may collude with each other on behalf of a state-level adversary.
Finally, as part of our relay selection algorithms (Section \ref{sec:system}),
we consider a probabilistic relay selection strategy that minimizes the amount 
of traffic that is observable by any single attacker over a period of time.
\section{Measuring Adversary Presence}\label{sec:measurement} 
In this section, we investigate the prevalence of the adversary described
in Section \ref{sec:background}. First, we detail how prediction of AS paths 
between a source and a destination is performed and how sets of potential 
attacking ASes are generated. Then we present the experimental methodology used
to make these measurements. Finally, we present the results of these
experiments.

\subsection{\edit{Predicting potential attacker ASes}}
Adversaries that can exploit asymmetric routing present a challenge to measuring 
their prevalence. The addition of potential attackers on the reverse-path between 
a source and destination implies the need for identifying potential attackers
(\ie ASes) on the reverse-paths between the client and entry-relay (and the 
exit-relay and destination). This poses a challenging measurement problem, since 
reliably measuring information about reverse-paths is currently not possible. 
While Reverse Traceroute \cite{reverse-traceroute} would be a useful tool for 
these measurements, it is currently not widely deployed.

Additionally, since our measurement toolkit was assembled with the goal of
integration with our Tor client -- \systemname (Section \ref{sec:system}), using
external measurement and control-plane mapping tools was not an option. This is 
because such tools require knowledge of the clients' intended destination -- an 
undesirable option for an anonymity tool such as Tor. Thus, any measurement or 
path prediction needs to be performed on the Tor client without leaking any 
information to attackers or third party tools and service providers.

To address the challenges of reliably measuring reverse-paths or use 
control-plane mapping tools, we employ an efficient path prediction approach 
which leverages up-to-date maps of the AS-level Internet topology 
\cite{Giotsas-IMC14}, and algorithmic simulations that take into account a 
common model of routing policies\cite{Gill-CCR12}. 


\myparab{AS-level topology. }We perform path prediction using an 
empirically-derived AS-level Internet topology. In this abstraction, the 
Internet is represented as a graph with ASes as nodes and edges as connections
between them. Connections between ASes are negotiated as business arrangements 
and are often
modeled as two main types of relationship: \emph{customer-provider} where the
customer pays the provider for data sent and received; and \emph{settlement-free
peering} or \emph{peer-peer} where two ASes agree to transit traffic at no cost
\cite{GR}.

However, in practice AS relationships may violate this simple taxonomy \eg ASes
that agree to provide transit for a subset of prefixes (\emph{partial transit})
or ASes that have different economic arrangements in different geographic
regions (\emph{hybrid relationships})~\cite{Giotsas-IMC14}. It can also be the
case that two ASes are controlled by the same organization \eg because of
corporate mergers such as Level 3 (AS3356) and Global Crossing (AS3549) or
organizations that leverage different AS numbers in different regions such as
Verizon (AS701, 702, 703). Additionally, integrating IXPs is a complicated
research subject due to a dearth of measurement data to inform how they should
be incorporated -- \eg just because two ISPs peer at an IXP does not mean all
paths including these ISPs will traverse the IXP. The AS-level topology we 
leverage takes partial transit and hybrid relationships into account, but
ignores IXPs (which would result in a significant over-estimation of our
measurements, due to their peering meshes). We use techniques discussed
and validated by Anwar \etal \cite{Anwar-TR15} for detecting sibling ASes. 
This is done to identify ASes that are likely to collude with each other.

\myparab{Routing policies. }Routing on the AS-graph deviates from simple
shortest path routing because ASes route their traffic based on economic 
considerations. We use a standard model of routing policies proposed by Gao and 
Rexford \cite{GR}. The path selection process can be broken down into the 
following ordered steps: 

\begin{itemize} 

\item \textit{Local Preference (LP).} Paths are ranked based on their next hop:
customer is chosen over peer which is chosen over provider.  

\item \textit{Shortest Paths (SP).} Among the paths with the highest local
preference, prefer the shortest ones.

\item \textit{Tie Break (TB).} If there are multiple such paths, node $a$ breaks
ties: if $b$ is the next hop on the path, choose the path where hash, $H(a,b)$
is the lowest.\footnote{In practice, this is done using the distance between
routers and router IDs. Since we do not incorporate this information in our
model we use a randomized tie break which prevents certain ASes from ``always
winning''.} 

\end{itemize} 

This standard model of local preference \cite{GR} captures the idea that an AS
has incentives to prefer routing through a customer (that pays \edit{it}) over a
peer (no money is exchanged) over a provider (that it must pay).

In addition to selecting paths, ASes must determine which paths they will
announce to other ASes based on export policies. The standard model of export
policies captures the idea that an AS will only load its network with transit
traffic if its customer pays it to do so \cite{GR}: 

\begin{itemize}

\item \textit{Export Policy (EP).} AS $b$ announces a path via AS $c$ to AS $a$
iff at least one of $a$ and $c$ are its customers.

\end{itemize}

Computing paths following these policies using simulation platforms (\eg CBGP 
\cite{CBGP}) can be computationally expensive which limits the scale of 
analysis. Thus, we employ an algorithmic approach \cite{Gill-CCR12} that allows 
us to compute all paths to a given destination in $\mathcal{O}(|V| +|E|)$ where 
$|V|$ is the number of ASes and $|E|$ is the number of edges.

\begin{figure}[t] \centering
\includegraphics[width=0.35\textwidth]{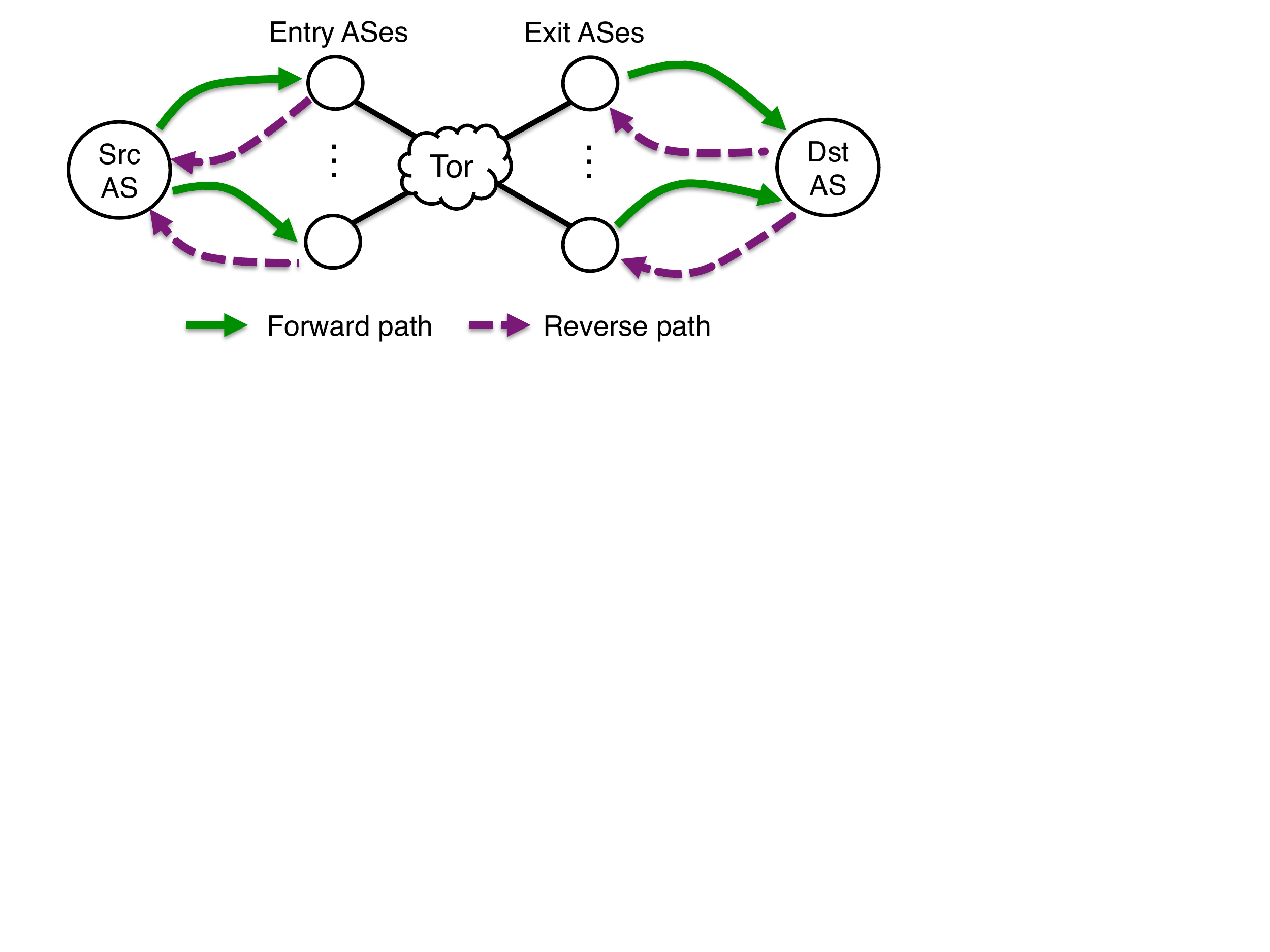}
\caption{Illustration of the AS paths that the client needs to predict, note
that these paths must be predicted for each potential entry and exit relay in
both the forward and reverse direction.} 
\label{fig:pathpredict} 
\end{figure}

\myparab{Predicting paths. }We use the routing policies and algorithmic
simulations \cite{Gill-CCR12} as described above to compute routes between pairs
of ASes using the AS-level topology published by CAIDA~\cite{Giotsas-IMC14}.
AS-level path prediction between a source and destination is a thorny 
issue, for example the recent work from Juen, \etal\cite{juen-pets15} shows 
that the paths predicted by BGP-based path prediction vary significantly 
from traceroute-based path prediction. However, our BGP-based path prediction 
toolkit makes use of the state-of-the-art in path inference and AS-relationship 
inference that have both been extensively validated with empirical measurements 
by Anwar \etal\cite{Anwar-TR15} and Giotsas \etal\cite{Giotsas-IMC14}.

In particular, Anwar, \etal\cite{Anwar-TR15} show that 65-85\% of measured paths 
are in the set of paths which satisfy \emph{LP} and \emph{SP}. Thus, we
modify the algorithmic simulator to return all paths satisfying \emph{LP} and
\emph{SP} simultaneously, instead of using \emph{TB} to produce a unique path.
Thus we consider the set of ASes in the set of paths satisfying \emph{LP} and
\emph{SP} between $a$ and $b$ to be the set $p_{a \rightarrow b}$.

%
%
%
%
\begin{figure}[t]
\centering
\includegraphics[trim=0cm 0cm 0cm 2cm, clip=true, width=0.45\textwidth]
{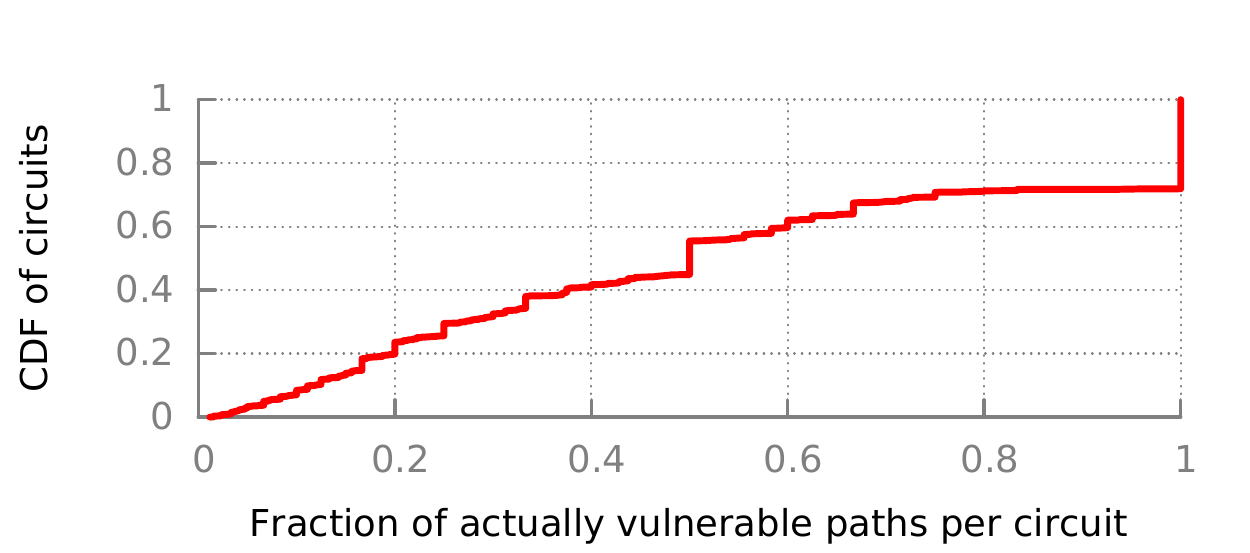}
\caption{\edit{Fraction of actually vulnerable paths from all possible paths,
for each of 20,000 circuits marked as vulnerable by our toolkit.}}
\label{fig:lower-bound}
\end{figure}

\myparab{Identifying vulnerable circuits. }
\edit{
Let $p_{src \leftrightarrow entry}^{i}$ be the $i^{th}$ \emph{LP} and
\emph{SP} satisfying (forward- or reverse-) path between the source and
entry-relay, $p_{exit \leftrightarrow dst}^{j}$ be the $j^{th}$ such path
between the exit and destination,
$\mathcal{P}_{src \leftrightarrow entry} = \cup_{i} \{p_{src \leftrightarrow
entry}^{i}$\}
, and 
$\mathcal{P}_{exit \leftrightarrow dst} = \cup_{j} \{p_{exit \leftrightarrow
dst}^{j}\}$. We refer to $\mathcal{P}_{a \leftrightarrow b}$ as the path-set
between $a$ and $b$.} 

\edit{Since it is currently not possible to predict exactly which path from
$\mathcal{P} = \mathcal{P}_{src \leftrightarrow entry} \times \mathcal{P}_{exit \leftrightarrow
dst}$ will be utilized when using a circuit with entry-relay $entry$ and
exit-relay $exit$, we label all paths $p \in \mathcal{P}$ as
vulnerable \emph{iff} at-least one of the paths in $\mathcal{P}$ is vulnerable 
(as defined in Eq. \ref{eq:attacked}). That is, once our path prediction 
toolkit returns the set of ASes that occupy each
path-set between the Tor client and a given entry-relay
$(\mathcal{P}_{src \leftrightarrow entry})$ and between the exit-relay and
destination $(\mathcal{P}_{exit \leftrightarrow dst})$,
potential circuits using the corresponding entry- and exit-relay
are labeled as vulnerable \emph{iff} there are
common or sibling ASes on the (client, entry-relay) and (exit-relay,
destination) path-set -- i.e., $\{\mathcal{P}_{src \leftrightarrow entry} \cap
\mathcal{P}_{exit \leftrightarrow dst}\}\ne \emptyset$.
This provides an estimate on the threat posed by network-level attackers.}

\edit{To understand the tightness of this estimate, we analyzed the fraction
of the actually vulnerable paths in each of 20,000 unique ``vulnerable'' 
circuits generated by our experiments. Figure \ref{fig:lower-bound} shows the 
result of this analysis. 25\% of all circuits had all their paths in
$\mathcal{P}$ vulnerable
to at-least one network-level attacker and 56\% of all circuits had at-least
50\% of their paths (in $\mathcal{P}$) vulnerable to at-least one network-level attacker.}
%
\begin{table*}[!t]
\begin{tabularx}{.95\textwidth}{|l|X|c|p{1.5in}|c|}
\hline
\textbf{ID} & \textbf{Question Answered} & \textbf{Vantage Point}
& \textbf{Setting} & \textbf{Results}\\\hline
E1 & 
How vulnerable are circuits to asymmetric correlation attacks? & 
 VPN & 
Live (3 guards) & Figures \ref{fig:vanilla-tor-country}, \ref{fig:any-as-level}
and \ref{fig:main-as-level} 
\\\hline
E2 & 
How many \emph{attacker-free} paths are available to the vanilla Tor client in 
each country? & 
100 ASes per country & 
Simulation (all entry- and exit-relays) & Figures \ref{fig:ecdf-sim} and
\ref{fig:five-percent}\\\hline
E3 &
How much of a threat do colluding sibling ASes pose? &
 VPN &
Live (3 guards) & Figures \ref{fig:vanilla-tor-siblings},
\ref{fig:any-colluding}, and \ref{fig:main-colluding}
\\\hline
E4 &
How much of a threat do state-level attackers pose? &
 VPN &
Live (3 guards) & Figures \ref{fig:vanilla-tor-state}, \ref{fig:any-state}, and
\ref{fig:main-state}
\\\hline
E5 &
Do guard settings have a significant effect on the availability of attacker-free
paths to the vanilla Tor client? &
 100 ASes per country &
Simulation (20 guard-sets of 1,2, and 3 guards and all exit-relays) &
Figure \ref{fig:guards-results}
\\\hline
\end{tabularx}
\caption{\edit{Summary of security experiment settings used for the evaluation of the
vanilla Tor client and \systemname. For each country, all experiments used a dataset 
containing the local Alexa Top 100 and 100 {locally sensitive}
websites (obtained from the Citizen Lab testing repository
\cite{citizenlab-blocked}). }
\label{tab:security-experiments}}
\end{table*}

\subsection{Measurement methodology and results}\label{subsec:measurement-results} 

To understand the threat posed by the adversary described in Section
\ref{sec:background}, we performed several experiments. In particular,
our goal was to understand the threat faced by the Tor client under various
configurations, and in different network and geographic locations.

\myparab{Experimental setup. }
In our experiments, we consider the fact that Tor users in different countries
face different levels of threats from local ASes. To this end, each experiment
was performed in 10 different countries: Brazil (BR), China (CN), Germany (DE),
Spain (ES), France (FR), England (GB), Iran (IR), Italy (IT), Russia (RU),
and the United States (US). This list was obtained by considering 
the intersections of the number of Tor users in each country \cite{tor-metrics} 
and the Freedom House rankings for Internet freedom \cite{freedom-house}. In
order to completely understand the threats faced by Tor users, five experiments 
were conducted in each country; a summary of each experiment is shown in Table 
\ref{tab:security-experiments}.

For each experiment, 200 websites were loaded using the Selenium Firefox
webdriver \cite{selenium}. The list of 200 websites comprised of the local 
Alexa Top 100 sites \cite{alexa-top} and 100 sensitive (i.e., likely to be
blocked) pages obtained from the Citizen Lab testing list repository 
\cite{citizenlab-blocked} for each country.

Each experiment was conducted in one of two settings: Live or Simulation. In
the Live setting, the actual client (vanilla Tor or Astoria) being studied 
was used to load pages from within the respective country using a single VPN as 
the vantage point. The VPN vantage point only presents a limited
picture of the threat faced by all users in the country (since it only considers
a single AS as the client location (source AS)), thus we used simulations to 
augment the Live experiments. Each simulation considered clients located in 100 
randomly selected ASes in each country.

For each experiment, logs were maintained to track: (1) the list of available
entry- and exit-relays during circuit construction, (2) the actual chosen entry 
and exit-relay for each circuit constructed by the client, and (3) the list of
requests made for each site and the circuit used by the Tor client to serve the
request. Data from these logs were fed to our measurement toolkit in order to
identify (1) the set of attackers that threaten actually constructed circuits 
(Live experiments) and (2) the set of attackers that threaten potential circuits 
-- i.e., circuits that could have been constructed given a particular valid 
combination of available entry- and exit-relays (Simulation experiments).
%
%
%
%

\myparab{E1: Measuring vulnerability to network-level attacks. }This
experiment was conducted using the vanilla Tor client and a modified Tor client
using a uniform relay-selection strategy. Both clients used the same VPN in
each of the 10 countries to load their corresponding Alexa top 100 and 100 
sensitive pages. Three statistics were measured: (1) The number of
websites which had the circuits carrying the request for their main page being
vulnerable, (2) the number of websites which had any of their circuits being
vulnerable, and (3) the total number of vulnerable circuits. 

A summary of these results are illustrated in Table 
\ref{tab:measurement-results-summary}. We see that both clients have similar 
number of compromisable circuits, however the vanilla Tor client allows 16\% 
more websites to load without having any of their circuits compromised, 
implying that when a website is loaded with the vanilla Tor client it is either 
completely safe or has most of its content loaded via a vulnerable circuit. 
This is due to the fact that unlike the modified Tor client, the vanilla 
Tor client reuses a small number of circuits for many requests.

\begin{table}  
\begin{minipage}{.45\textwidth}
\begin{center}
\begin{small}
\begin{tabularx}{\textwidth}{|l|X|X|}
\hline
\textbf{Vulnerable}	    & \textbf{Vanilla Tor} 	  & \textbf{Uniform Tor}       \\\hline
\textbf{Websites (Main request)} &  37\%     & 35\%   \\\hline
\textbf{Websites (Any request)}  &  53\%     & 69\%   \\\hline
\textbf{Circuits (All requests)}  &  40\%     & 39\%   \\\hline
\end{tabularx}
\end{small}
\end{center}
\caption{\edit{Summary of threat from asymmetric correlation attacks against the 
vanilla Tor and uniform relay-selection strategies for 200 websites in 10 
countries. \label{tab:measurement-results-summary}}}
\end{minipage}
\end{table}

\begin{figure}[t]
\centering
\includegraphics[trim=0cm .5cm 0cm 2cm, clip=true, width=0.45\textwidth]
{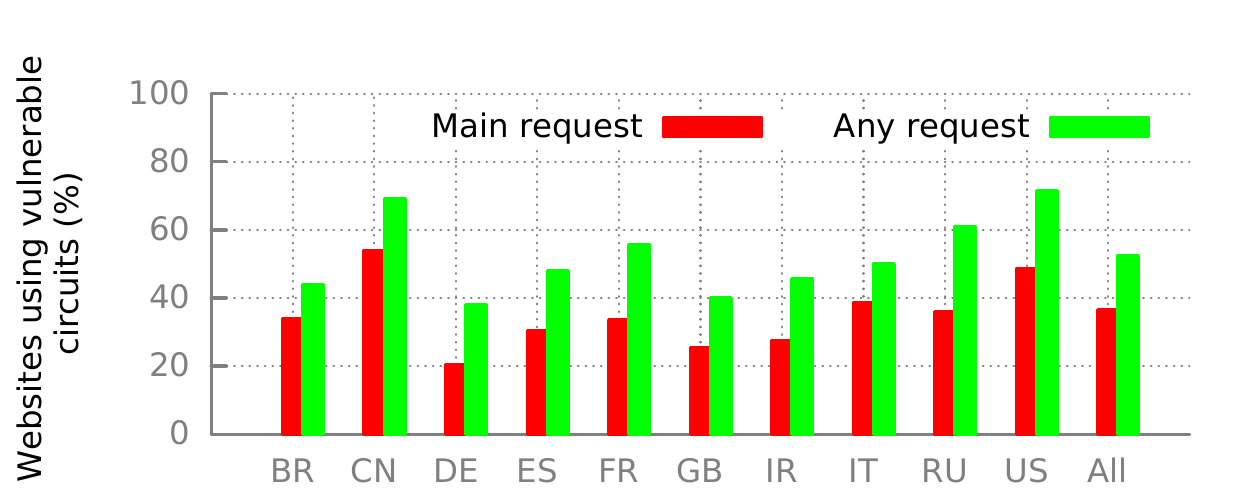}
\caption{\edit{An estimate of the percentage of websites that have main page requests and any
requests serviced by a vulnerable Tor circuit.}}\label{fig:vanilla-tor-country}  
\end{figure}

We break down our results for the vanilla Tor client by country in Figure 
\ref{fig:vanilla-tor-country}. The figure shows the percentage of websites that 
are vulnerable to asymmetric correlation attacks on circuits built for serving 
the request for their main page (GET) and for serving any request. 
We find that the threat is not uniformly spread. Clients using
the vanilla Tor client from our VPN vantage point in three countries: China
(CN), Russia (RU), and the United States (US) were found to be most vulnerable.
This can be explained by the fact that of our 10 countries, the US, RU, and CN
had the most amount of locally hosted content (i.e., content hosted within the
country). Of the 200 sites used for each of the countries, 95\% (US), 57\%
(RU), and 47\% (CN) made requests to ASes within the country itself -- making 
it more likely for  the same AS to be on paths from/to client to/from 
entry-relay and exit-relay to/from destination.

\myparab{E2: Measuring fraction of \emph{available attacker-free} paths.} 
Since the results of our experiments on the live Tor network were highly 
dependent on the location of the VPN, simulations were required to understand 
the distribution of threat in other locations within each country. To this end, 
for each country, 100 ASes were randomly selected as client locations and the 
targets of the each of the requests generated by the 200 sites (sensitive and
popular) for each of our 10 countries were used as destinations. The simulation 
toolkit generated a list of all entry- and exit-relays available to each client 
for performing the page load (using Tor client consensus data).

\begin{figure*}[htb]
\centering
\begin{subfigure}[b]{0.495\textwidth}
\includegraphics[trim=0cm 0cm 0cm 0cm, clip=true,width=\textwidth]
{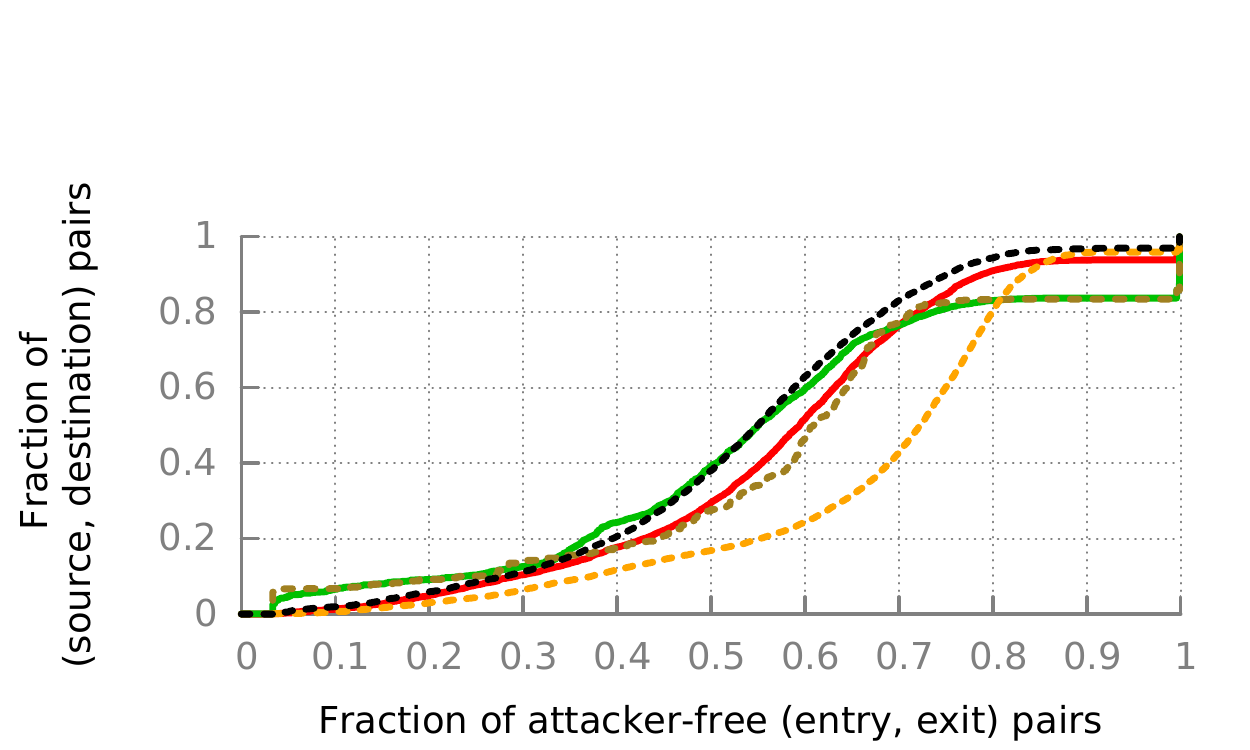}
\caption{Most vulnerable countries (all websites): BR, CN, IR, RU, US}
\label{fig:ecdf-top-5}
\end{subfigure}
\begin{subfigure}[b]{0.495\textwidth}
\includegraphics[trim=0cm 0cm 0cm 0cm, clip=true,width=\textwidth]
{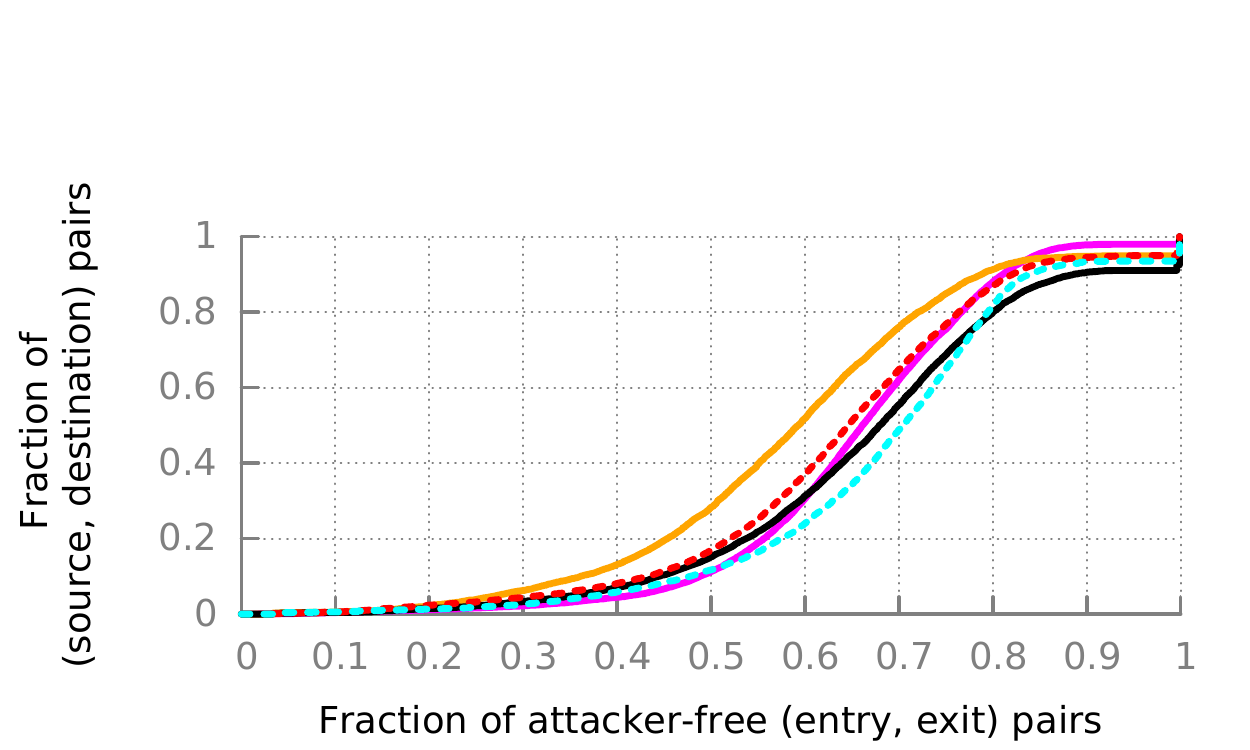}
\caption{Least vulnerable countries (all websites): DE, ES, FR, GB, IT}
\label{fig:ecdf-bottom}
\end{subfigure}
\begin{subfigure}[b]{0.495\textwidth}
\vspace{-.1in}
\includegraphics[trim=0cm 2cm 0cm 5cm, clip=true,width=\textwidth]
{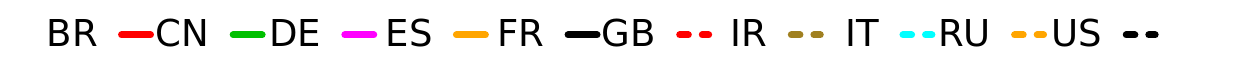}
\end{subfigure}
\caption{\edit{Distribution of the fraction of attacker-free 
circuits for 100 source ASes connecting to 200 websites in 10 different countries of interest. More skewed to the
right indicates the availability of more safe circuits.}}
\label{fig:ecdf-sim}
\end{figure*}

Each generated (source, entry, exit, destination) combination  was then analyzed 
for the threat of attackers to understand how many ``safe'' or ``attacker-free'' 
entry-exit pairs were available. We see in Figure \ref{fig:ecdf-sim} the cumulative
distribution function of the fraction of attacker free entry-exit pairs for each
source-destination pair. Figure
\ref{fig:ecdf-top-5} shows this for the five most vulnerable countries in our
study, and \ref{fig:ecdf-bottom} shows this for the remaining countries.

China (CN) and Iran (IR) stand out as the most interesting cases. First, we see 
that 8\% of all source-destination pairs have less than 
10\% of their entry-exit options being safe. Next, we also 
notice that there are no known attackers present on 18\% of all
source-destination pairs. This 
appears to indicate that the threat of de-anonymization is non-uniform even 
within a country, with certain client locations being much safer than others.

\begin{figure}[htb]
\centering
\includegraphics[trim=0cm .42cm 0cm 2cm, clip=true,width=.45\textwidth]
{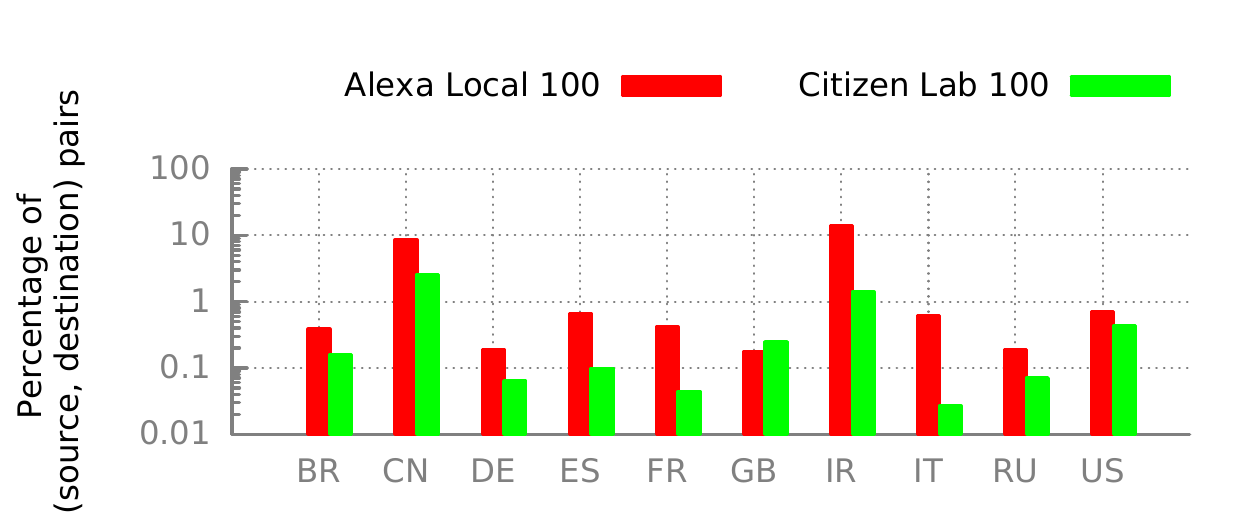}
\caption{\edit{(\textbf{Logscale}) Percentage of (source, destination) pairs having fewer than 5\% attacker-free (entry, exit) 
options in each country.}}
\label{fig:five-percent}
\end{figure}

In order to understand which set of websites are more vulnerable in each of the
countries, in Figure \ref{fig:five-percent} we show the percentage of source-
destination pairs having fewer than 5\% safe circuit options for each set of
websites. We find that in
all cases, the Alexa top 100 local websites have fewer safe circuit options.
This can be explained by the fact that locally popular websites are likely to be
hosted within a regional AS. Additionally, we find that China and Iran have a
significant number of their source-destination pairs having fewer than 5\% safe
circuit options -- i.e., over 8\% of the source-destination pairs have less than
5\% of all their circuit options being safe from network-level correlation
attacks.

However, in general, the results of \textbf{E1} and \textbf{E2} indicate that although in most cases
there are many safe entry-exit options available to the Tor client, it often 
does not select these options -- leading to a large number of vulnerable 
circuits being created.

%
%
\myparab{E3: Measuring the impact of sibling ASes. }In this experiment we
consider the possibility that ASes owned by the same organization (referred to
as sibling ASes) may collude with each other in order to de-anonymize Tor users 
via asymmetric correlation attacks. We use data gathered by Anwar \etal
\cite{Anwar-TR15} to identify such ASes. The same setup as \textbf{E1} was used.

\begin{figure}[t]
\centering
\includegraphics[trim=0cm .5cm 0cm 2cm, clip=true, width=0.45\textwidth]
{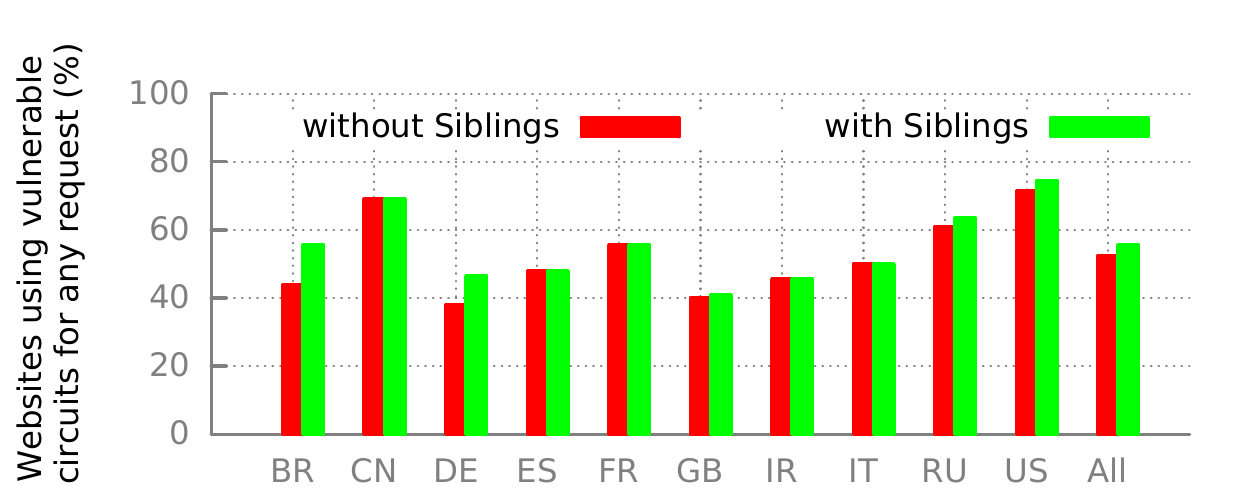}
\caption{\edit{An estimate of the percentage of websites that have any requests 
served by a vulnerable Tor circuit when considering siblings.}}
\label{fig:vanilla-tor-siblings}
\end{figure}

We observe from Figure \ref{fig:vanilla-tor-siblings} that the increase in 
threat from considering sibling ASes is marginal. Over the 10 countries, 
only 3\% additional websites from our list of 200 for each country had some 
request served by a circuit that was vulnerable to asymmetric attacks by 
sibling ASes. However, the increase in threat is not uniform. Clients in 
Brazil and Germany face an 8-10\% increase in vulnerable websites. This can be
attributed to the large telecom conglomerates operating within the countries --
\eg many paths from our vantage points in Germany and Brazil were vulnerable to 
correlation attacks due to transiting one of the large number of ASes owned 
by Telefonica (in Spain) and Durand (in Brazil), respectively.

\myparab{E4: Measuring the impact of state-level adversaries. }In this
experiment we consider the threat that Tor clients face from state-level
adversaries. We assume that a state-level adversary is able to gain insight
into the traffic flowing through all ASes operating within the state.
Therefore, we consider a circuit originating from country X to be vulnerable if
its path to/from its entry-relay and from/to the exit-relay to the destination 
contains some AS operating within X. The same setup as \textbf{E1} 
was used for data collection.

\begin{figure}[th]
\centering
\includegraphics[trim=0cm .05cm 0cm 2cm, clip=true, width=0.45\textwidth]
{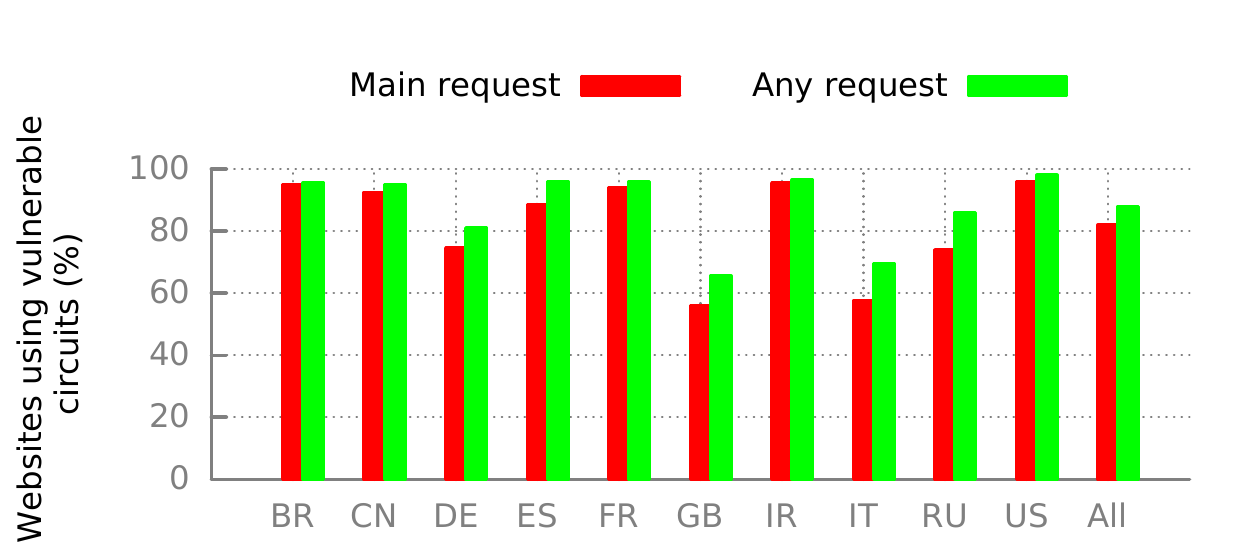}
\caption{\edit{An estimate of the percentage of websites that have main page
requests or any 
requests served by a vulnerable Tor circuit when considering state-level 
adversaries.}}
\label{fig:vanilla-tor-state}
\end{figure}

The results are broken down per country in Figure \ref{fig:vanilla-tor-state}.
Here, we see that the situation is quite dire with $82\%$ of all (over
all 10 countries) websites having their main page served by a vulnerable 
circuit. In particular, clients in Brazil, China, France, Iran, and the United 
States face the biggest threat from state-level attacks
with over 95\% of their main page requests being vulnerable to state-level
attackers.

\myparab{E5: Measuring the effect of guards. }In this experiment we
consider the effect of the number of guards on the vulnerability of Tor clients
to network-level asymmetric correlation attacks. For each of our 10 countries,
100 ASes were randomly selected as client locations and the targets of all the
requests generated by the 200 websites in our earlier experiments were used as
the destinations. The simulation toolkit generated 60 unique guard-sets (20
each for 3 guards, 2 guards, and 1 guard) in an identical manner to the
vanilla Tor client, and a list of all exit-relays available
to each client for performing the page load (using Tor consensus data). Each
(source, entry, exit, destination) combination was checked for the presence of our
adversary. 

\begin{figure}[th]
\centering
\includegraphics[trim=0cm 0cm 0cm 2cm, clip=true, width=0.45\textwidth]
{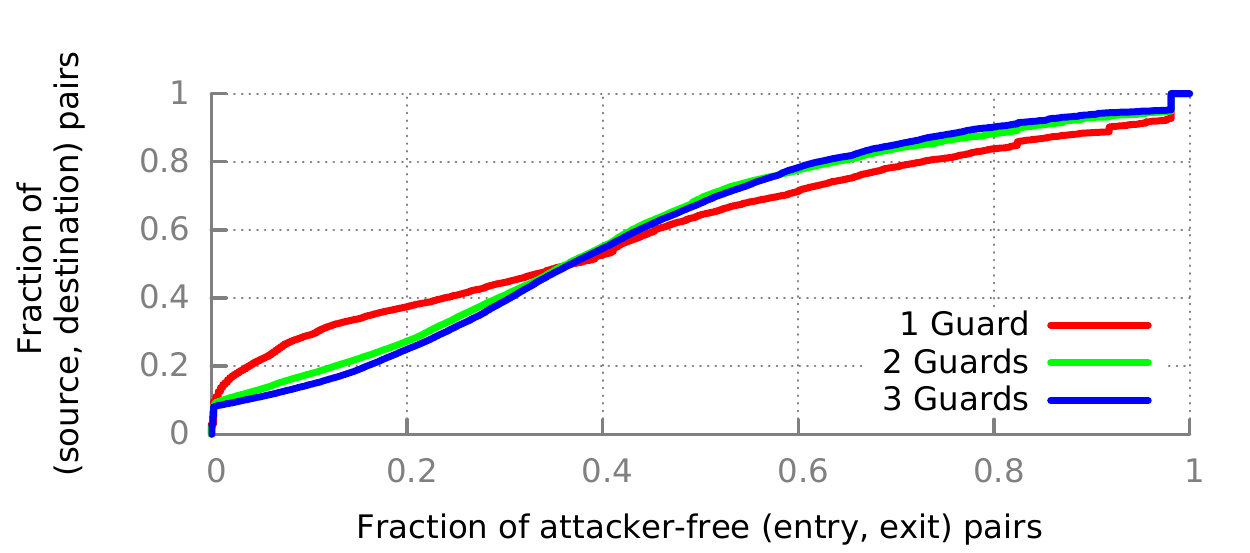}
\caption{\edit{Distribution of the fraction of attacker-free (entry, exit) pairs for
vanilla Tor with 3, 2, and 1 guard(s).}}
\label{fig:guards-results}
\end{figure}

Figure \ref{fig:guards-results} illustrates the effect that reducing the size
of the guard-set has on the fraction of network-level attacker-free-paths 
available to the Tor client.

While it is known that a smaller number of guards provides better security
against relay-level attackers in the long-term \cite{guards}, we see from the
results of this experiment that the effect is the opposite against network-level 
adversaries -- i.e., as the size of the guard-set decreases, Tor is more likely 
to select a circuit vulnerable to network-level asymmetric correlation attacks 
due to the reduced number of available safe paths. In particular, when only 1
guard is used, over 15\% of the (source, destination) pairs in our experiment 
had no safe-options, whereas the difference in security provided by two or three
guards was marginal. This experiment demonstrates one of the conflicts between 
Tor clients geared for defending against relay-level attackers and those geared 
for defending against network-level attackers.

\section{\systemname: An AS- and Capacity-aware Tor Client}\label{sec:system}

Motivated by the observation that vanilla Tor very often selects entry-exit
pairs that may be subject to asymmetric correlation attacks, we seek to 
design a relay selection algorithm to mitigate the opportunities for such 
attackers. We design our relay selection system, \systemname,  based on the idea 
of stochastic relay selection. This works by having the Tor client generate a 
probability distribution that minimizes the chance of attack over all possible
entry- and exit- relay selection choices, and selecting an entry- and exit-relay 
based on this 
distribution. The advantage of stochastic selection is that even if the 
client has no safe options, relay-selection can be engineered to minimize the 
amount of information gained by the adversary over some period of time (as we 
show below). Further, it allows clients to select relays in a way such that no set 
of relays in the Tor eco-system is overloaded, even if every client uses the 
same relay-selection strategy. 

\subsection{\systemname goals}

\systemname is constructed with several security and performance goals in mind:

\begin{itemize}

\item \emph{Deal with asymmetric attackers. }\systemname avoids constructing
circuits involving common ASes on the forward- or reverse-paths between the 
client to the entry-relay and the exit-relay and the destination. 

\item \emph{Deal with the possibility of colluding attackers. }\systemname
considers the threat of ASes that may collude to de-anonymize
Tor users. \systemname can be configured to build circuits that
do not contain known to be colluding ASes on the forward- or reverse-path
between the client and entry-relay and exit-relay and destination. This
mitigates the threat from sibling ASes and state-level attackers.

\item \emph{Consider the worst case possibility. }\systemname uses a
probabilistic relay selection algorithm that ensures, even in the worst-case
(where there are no safe paths to and from the entry- and exit-relay), that
the ability of a single AS (or, family of ASes) to de-anonymize a large number 
of circuits is minimized.
 
\item \emph{Minimize performance impact. }It is clear that any AS-aware client
will lose its ability to perform 
many optimizations such as pre-constructing circuits. Our goal is to minimize
the effect of the above considerations  on the performance of the Tor client.

\item \emph{Be a good network citizen. }\systemname takes into account the
capacities of all relays available in the Tor eco-system and performs selection
in a way that no single set of relays are overloaded, even when all clients in
the network use the same relay-selection strategy.

\end{itemize} 

\subsection{Minimizing information gained by the adversary} \label{sec:LP}

While there often are cases when there is a relay selection that will completely 
eliminate the risk of our adversary, we develop our relay selection to be robust, 
even if this is not the case. Further, with attacks implemented using 
BGP hijacking and interception the number of unsafe paths may be higher than 
what we observe in our analysis (we discuss this more in Section 
\ref{sec:discussion}).

To minimize the risk of correlation attacks, we define a linear program which 
generates a probability for each relay selection with the objective to minimize 
the maximum probability of a circuit encountering the attacker. Recall that in 
our adversary model, we consider a long-lived adversary and that minimizing 
the probability of an attacker may also be seen as minimizing the number of 
circuits the adversary is able to observe over a long period of time and 
numerous circuit construction cycles. 

\begin{figure}[t] 
\centering
\includegraphics[width=0.4\textwidth]
{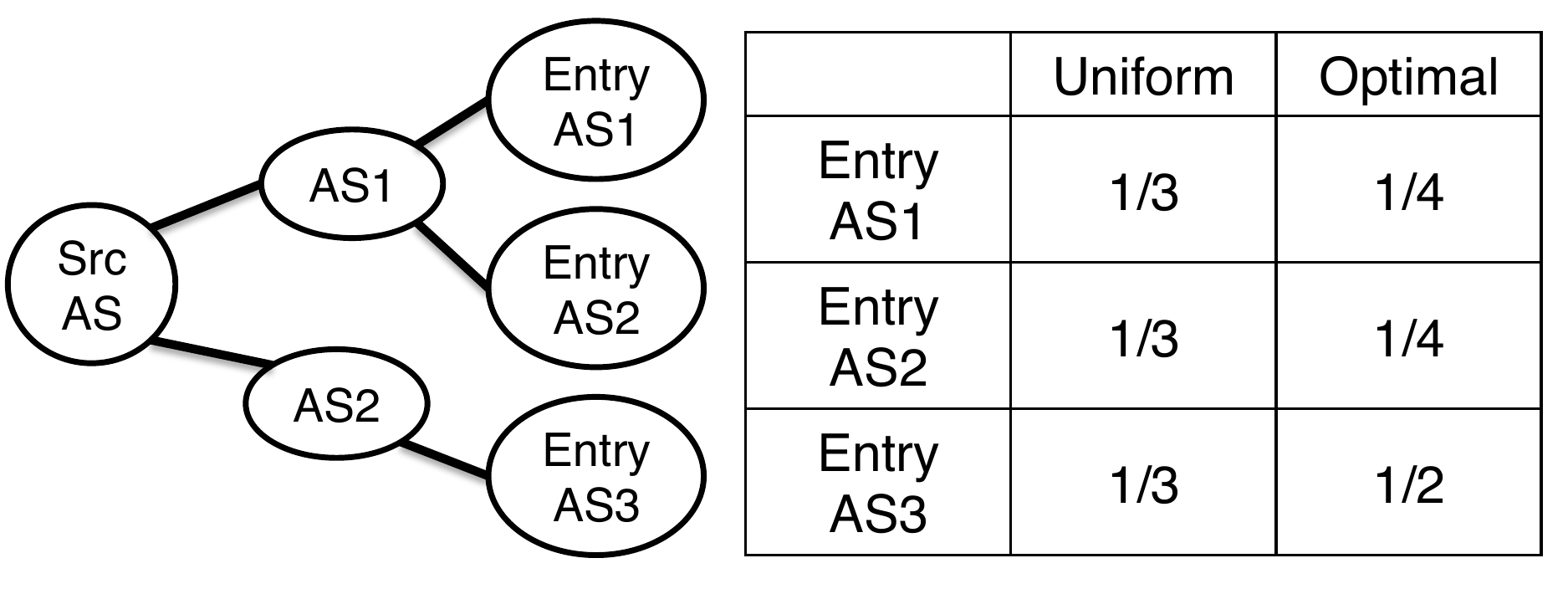} 
\caption{Example of optimizing relay selection. Simplified to unidirectional 
paths and only entry-relay selection.} 
\label{fig:lpexample} 
\end{figure}

Figure~\ref{fig:lpexample} shows an example of relay selection to give intuition
about how the LP minimizes the risk from the attacker. In this example, we
consider unidirectional paths and only entry-relay selection for clarity. In the 
figure, if the source were to choose uniformly at random across the three 
entry-relays, there is a 2/3 chance that $AS1$ will be able to observe traffic 
and only a 1/3 chance that $AS2$ will. In this case, the optimal selection is
intuitive, that the source should choose entry-relays 1 and 2 with probability
1/4 each and entry-relay 3 with probability 1/2. This lowers the probability
that $AS1$ can observe a circuit from 2/3 to 1/2. This probability of the most
likely adversary is the quantity that our LP minimizes. 

We use the following notation:

\begin{itemize}

\item Let $ADV_{i, j}$ be the set of attackers on the circuit using 
entry-relay $i$ and exit-relay $j$ to destination $dest$ -- i.e., 
$\forall A \in ADV_{i, j} : A \in \{p_{src \leftrightarrow entry_i} \cap
p_{exit_j \leftrightarrow dest}\}$. 

\item Let $X_{i, j, A}$ be an indicator random variable for attacker $A$ on 
the circuit using entry-relay $i$ and exit-relay $j$ -- i.e., 
$X_{i, j, A} = 1 \iff A \in ADV_{i, j}$, and 0 otherwise.

\item Let $P_{i, j}$ be the probability that a client builds a circuit using 
entry-relay $entry_i$ and exit-relay $exit_j$. 

\end{itemize}

\edit{The following linear program is used to minimize the probability of the most
likely attacker (i.e., the number of circuits visible to the attacker).}

\begin{equation}\label{eq:lp}
\begin{alignedat}{4}
\text{minimize }   & z  \\
\text{subject to } & z~ \geq ~\sum_{i,~j} (P_{i,j} \ X_{i,j,A}) & ~~\forall A
\in ADV_{i, j}\\
		   & P_{i,j} \in [0,~ 1] ~,~ \forall i, ~\forall j &;~ \sum_{i,~j}~P_{i,j} = 1  
\end{alignedat}
\end{equation}

Essentially, given information about the presence of attackers (network-level
or state-level) for each $p_{source \leftrightarrow i}$ and 
$p_{j \leftrightarrow dest}$ path, the linear program seeks to find the 
probability distribution ($P_{i, j}$) over available choices of entry- and 
exit-relays, for which the expected number of circuits visible to each attacker 
is minimized. Entry- and exit-relays are chosen according to this distribution
\edit{(defined as $D_{lp}$)} during circuit construction.

\subsection{Security is not enough}\label{subsec:load-balancing}

While our LP produces a relay selection distribution that minimizes the 
probability of success across all adversaries, it does not take into account the 
resources available at the selected relays. Given that Tor is a system run using 
community resources contributed by volunteers, load balancing users across these 
resources is important to ensure that they are used efficiently and no single 
relay or set of relays become overloaded. Figure~\ref{fig:astoria-load} 
shows a snapshot of the distribution of relay capacities available during the 
period of this study, for all relays in the Tor system and the relays selected
by a hypothetical perfect load-balancing Tor client -- i.e., one where each
relay serves exactly the amount of traffic that it can handle (assuming
identically sized requests). Here, we see that over 80\% of all Tor traffic 
should be routed through $\approx 35\%$ of all the relays in the Tor network 
for every relay to be operating within its advertised capacity.


In order to achieve load-balancing, we augment our relay-selection algorithm
with information about relay capacities from the latest Tor consensus during 
circuit construction. This is done as follows:

 \emph{When there are safe entry and exit combinations: }In this 
case, we select a safe combination according to the distribution of relay
capacities. For example, given a set of safe entry- and exit- relay combinations
$E$ = $\{(en_1, ex_1) \dots (en_n, ex_n)\}$ and the distribution of their
advertised capacities $D_{bw}$ = $\{en_1,$ $\dots, en_n, ex_1, \dots, ex_n\}$, we 
select a combination $(en_i, ex_i)$ with probability $P_i$ = 
$\frac{D(en_i) \times D(ex_i)}{\sum_{j=1}^n {D(en_j)} \times D(ex_j)}$. 

This
ensures that no single (entry- or exit-) relay is selected with probability 
higher than the ratio of its advertised capacity and the total advertised 
capacity of all safe (entry- or exit-) relays (just as is done by the vanilla
Tor client).

 \emph{When there are no safe entry- and exit-relay combinations: }In this
case, in order to correctly minimize the amount of information gained by the
adversary, we strictly obey the probability distribution output by our linear 
program described in the previous section. No attempt is made to balance loads
according to relay capacities. It is important to note that this is a fairly
infrequent case (as shown in experiment \textbf{E2} in Section 
\ref{sec:measurement}).

\subsection{Implementing \systemname}


The measurement toolkit described in Section \ref{sec:measurement} was
integrated with a modified Tor client, as follows.

\myparab{Integrating our path measurement toolkit with the Tor client. }
For standard measurement purposes, the
toolkit simply takes a source and destination address and returns the set of
ASes on the forward and reverse-path between the two.

However, in the context of integration with the \systemname client, it must
predict paths to and from each of the entry-relays for the client's AS, and 
paths from all exit-relays toward the destination AS (Figure 
\ref{fig:pathpredict}). This results in $ |En| + |Ex| + 2$ routing-tree 
computations where $|En|$ and $|Ex|$
are the number of entry and exit relays, respectively. In order to mitigate the
risk of correlation attacks, by default, Tor restricts the number of
entry-relays available to each Tor client to three (called guards
\cite{guards}), and there are typically of the order of 1,000 exit-relays
available to a client during circuit construction -- resulting in the order of
1,000 routing-tree computations.

Fortunately, since the source AS and entry-relay ASes are relatively stable,
these paths can be precomputed for later use by the client. (We observe the
benefit of this in Section~\ref{sec:evaluation}.) However, performing
relay selection on a per-destination basis means that pre-building circuits, as
is done by the current implementation of Tor, is no longer feasible. 

\myparab{AS-aware on demand circuits. }First, the Tor client was modified 
to perform offline IP to ASN mapping using a database 
\cite{apnic-asn-map} for every incoming request. Note that since the entire 
database (9 MB) is downloaded, the client does not reveal its intended 
destination to any lookup services. 

Next, modifications were made to the way requests were allocated to circuits. 
The vanilla Tor client performs pre-emptive circuit construction in order to 
serve requests as they arrive (increasing performance significantly). This is
unfortunately infeasible for a AS-aware client where relay-selection
is a function of the destination.
Although one may consider pre-constructing AS-aware circuits for a set of 
popular destination  ASes, the performance benefit is marginal, at best. 
This is mainly due to the large number of third party requests for less popular 
destination ASes embedded in popular Web pages. Astoria, therefore, only performs 
on demand circuit construction. For each incoming request, Astoria first checks 
if there are existing circuits serving the same destination AS. The request is 
attached to the most suitable such circuit if it exists. 

\myparab{Circuit construction. }Astoria creates a new 
circuit if and only if a request arrives for a destination with no currently
usable circuits. In such cases, the
client and destination ASNs are passed to the circuit construction and relay
selection algorithms. Circuit construction is performed as follows:

\begin{itemize}

\item First, a list of entry- and exit-relays meeting the requirements set by
the request were obtained. If the Tor client is configured to utilize only
guards as entry-relays, the list of guards is obtained. Next, in order to
perform load-balancing, information from the most recent Tor consensus is 
obtained to generate the relay capacity distribution $D_{bw}$ for each entry- 
and exit-relay combination.

\item The Astoria client performs lookups to the offline IP-ASN database to
perform mapping between entry- and exit-relay IP address and AS numbers. These,
along with the client and destination AS numbers are then passed to our AS-path 
prediction and attacker measurement toolkit (Section \ref{sec:measurement}).

\item The toolkit returns the list of ASes on each forward- and reverse-path
between the client and every potential entry-relay and the destination and every
potential exit-relay. In order to improve performance, paths are cached for
frequently queried destinations. Precomputation or caching of paths between the 
client and the high-uptime entry-relays and destinations and high-uptime 
exit-relays also help improve performance.

\item The returned paths are checked for the presence of common ASes in the
entry and exit AS path sets. If there are paths without an attacker, the linear
program need not be invoked. Instead, Astoria selects a safe entry- and 
exit-relay combination according to the generated $D_{bw}$ probability 
distribution (described in Section \ref{subsec:load-balancing}). We see the 
impact of this load-balancing technique in Section \ref{sec:evaluation}.

\item If there are no attacker-free relay combinations, the linear program is
invoked in order to select an entry- and exit-relay combination according to
the distribution $D_{lp}$ that minimizes the probability of the most likely 
attacker (described in Section \ref{sec:LP}).

\item Finally, once the entry- and exit-relays are selected according to one of
the $D_{bw}$ or $D_{lp}$ distributions, the circuit is constructed. The 
remainder of the circuit construction process remains unchanged from the
vanilla Tor client.
\end{itemize} 
 

\section{Astoria Evaluation}\label{sec:evaluation}

We evaluate \systemname along multiple axes. First, we consider the performance
of \systemname by measuring the time required to load webpages and its ability 
to be a good Tor citizen by selecting bandwidth-rich relays. Second, we evaluate 
the security provided by \systemname. We show that \systemname constructed 
circuits are a good defense against the adversary described in Section 
\ref{sec:background}. Finally, we evaluate the threat from attacks by
relay-level adversaries.

\subsection{Evaluation methodology}

Similar to our experiments in Section \ref{sec:measurement}, we consider the
performance and security of clients in 10 different countries -- Brazil (BR),
China (CN), Germany (DE), Spain (ES), France (FR), England (GB), Iran (IR),
Italy (IT), Russia (RU), and the United States (US). The same 200 webpages as 
before were used for page-loads within each country.

In order to understand the performance of \systemname and for comparison with 
the vanilla Tor client, three metrics were computed: (1) page-load
times\footnote{The Selenium \texttt{driver.get()} method was used to detect 
the end of page-loads.}, (2) distribution of selected relay bandwidths, and 
(3) overhead of path prediction. For each of these experiments we considered 
the same experimental settings as the vanilla Tor client in experiment 
\textbf{E1}. Logs were recorded to extract advertised capacities of all 
available relays and all relays selected by the \systemname and vanilla Tor 
clients, and time required for AS path computation by the \systemname client.

In order to assess the security of \systemname and for comparison with the
vanilla Tor client, experiments to measure security against network-level
(experiment \textbf{E1}), colluding network-level (experiment \textbf{E3}), 
and state-level (experiment \textbf{E4}) asymmetric correlation attackers were 
repeated using the \systemname client for page-loads in the same setting
(including using the same guard-set in each country) as the vanilla Tor client 
(Section \ref{sec:measurement}). For each experiment, three statistics were 
computed: (1) the fraction of websites whose main page requests were served by 
vulnerable circuits, (2) the fraction of websites that any request that was 
served by a vulnerable circuit, and (3) the total fraction of vulnerable 
circuits.

\subsection{Performance evaluation}
In this section, we evaluate the performance of \systemname using three metrics: 
(1) page-load times, (2) distribution of selected relay bandwidths, and (3) 
overhead of path prediction. 

\begin{figure}[t] 
\centering
\includegraphics[trim=0cm 0cm 0cm 2cm, clip=true, width=0.45\textwidth]
{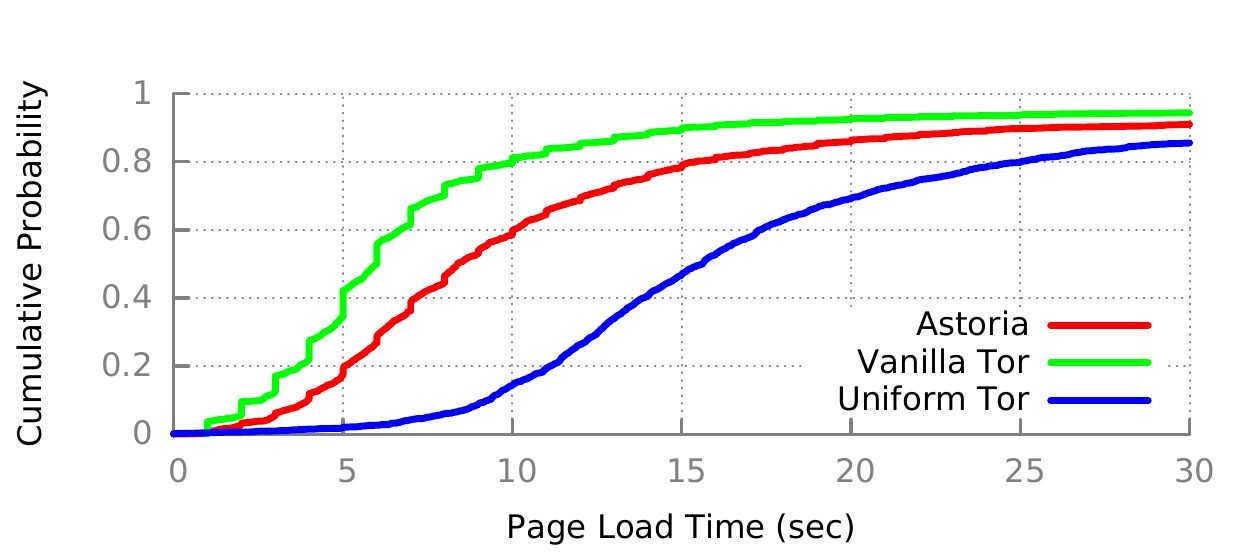}
\caption{\edit{CDF of page load times (including circuit creation times) for a uniform Tor, vanilla Tor, and \systemname 
client over 200 websites in all 10 countries.}}
\label{fig:page-load-times} 
\end{figure}

\myparab{Page load times. }Figure \ref{fig:page-load-times} shows the distribution of page-load times
when using the vanilla Tor client, a modified Tor client with a uniform
relay-selection strategy, and the \systemname client. We see that the median
page-load time with the vanilla Tor client is only \textbf{5.9} sec, while the
median page-load time for the \systemname and uniform Tor client are
\textbf{8.3} sec and \textbf{15.6} sec, respectively. Although this drop in
performance from the vanilla Tor client to \systemname is significant, it can 
be argued there are two main causes for this, both of which are unavoidable to
any AS-aware Tor client: (1) It is no longer possible to pre-construct and
re-use circuits to the same degree as the vanilla Tor client, and (2) There is
a non-negligible amount of time spent for computing paths and checking for the
presence of attackers on these paths.

\begin{figure}[t] 
\centering
\includegraphics[trim=0cm 0cm 0cm 2cm, clip=true, width=0.45\textwidth]
{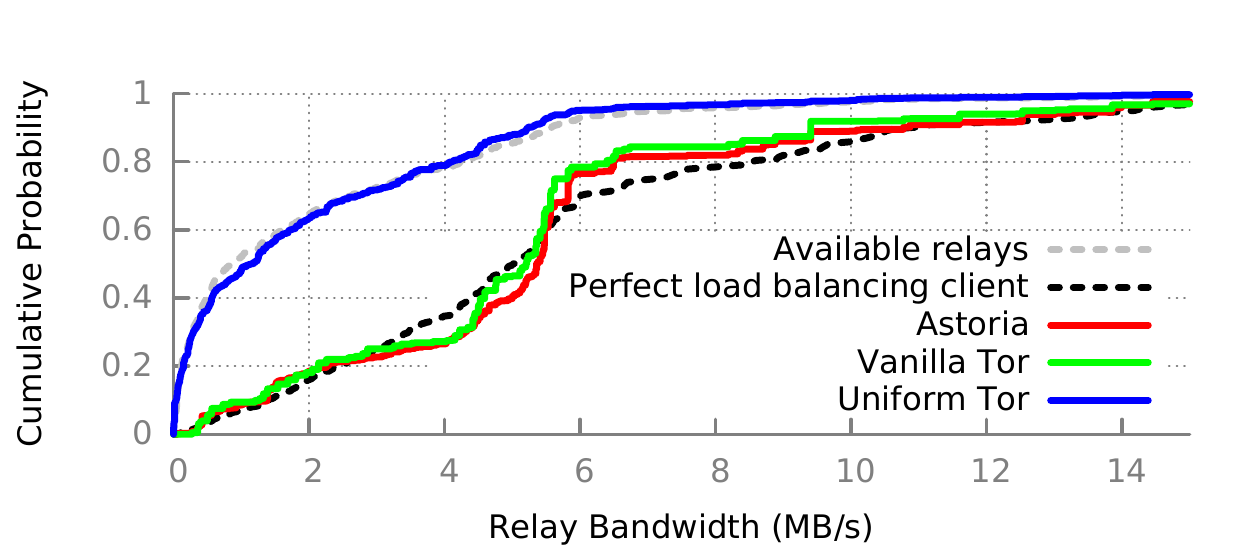}
\caption{Distribution of bandwidths of relays selected by vanilla Tor, uniform
Tor, \systemname, and the perfect load balancing client.}
\label{fig:astoria-load} 
\end{figure}

\myparab{Load balancing. }\systemname aims to balance load from clients across 
all relays in the Tor network so that no single set of relays are overloaded.
Figure \ref{fig:astoria-load} demonstrates the closeness of the load-balancing
of the \systemname client with the vanilla Tor client and the perfect
load-balancing client. We see that in spite of performing AS-aware
relay-selection, \systemname is able to perform load-balancing at least as well
as the vanilla Tor client, with neither of them achieving a perfect
distribution. 

The results of this experiment allow us to confirm our hypothesis that the 
reduction in performance from the vanilla Tor client to \systemname is indeed 
because of our inability to preconstruct circuits and delays due to path 
computation, and not due to poor relay-selection.

\begin{figure}[t]
\centering
\includegraphics[trim=0cm 0cm 0cm 2cm, clip=true,width=0.45\textwidth]
{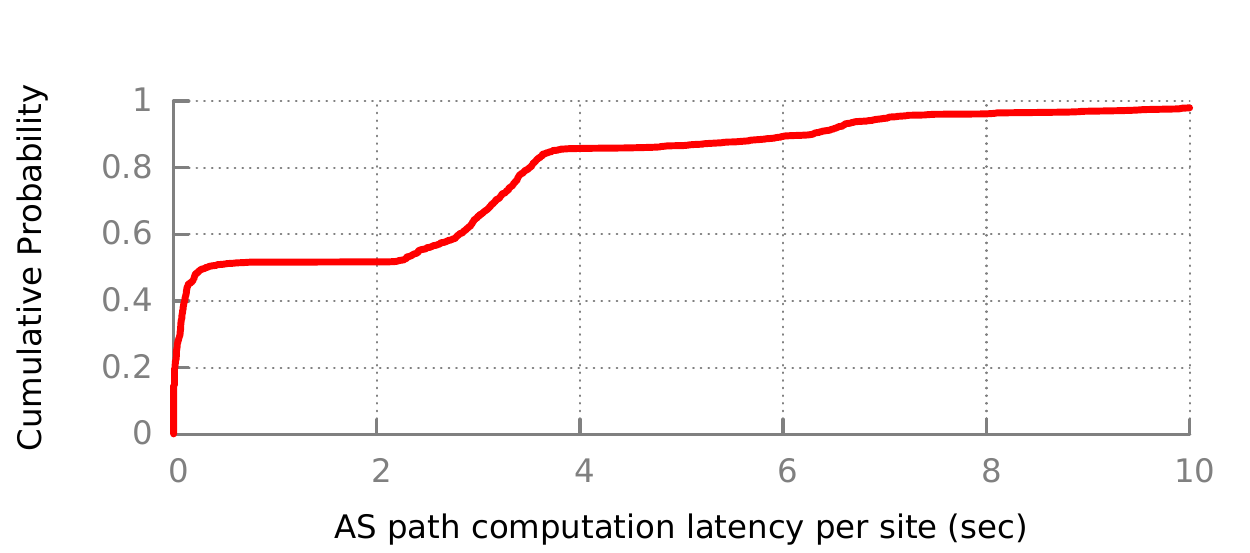}
\caption{CDF of time spent on AS path computation per site.}
\label{fig:bgp-sim}
\end{figure}

\myparab{Overhead of path prediction. }Figure \ref{fig:bgp-sim} shows the CDF of
the total amount of time spent on computing AS paths, for each site. We see that 
for about 50\% of all sites (200 sites in each of 10 countries), the time spent 
on path computation is negligible. This is due to the high frequency of repeated 
occurrences of destination ASes in our 200 sites -- resulting in the AS path 
for each exit-relay to that destination already being in the toolkit's cache. 
In 60\% of the cases where responses were not cached (and $86\%$ of the 
cases, overall), computing AS paths required under 4 seconds. 

\subsection{Security against network-level attackers}
In this section, \systemname is evaluated and compared with the vanilla Tor
client by measuring its success in defending against various attackers 
performing asymmetric correlation attacks. A summary of all results are
provided in Table \ref{fig:astoria-security}.

\begin{table*}
\renewcommand{\arraystretch}{1.2}
\centering
\begin{tabularx}{\textwidth}{|p{1.1cm}|X|X|X|X|X|X|X|X|X|}
\hline
\multirow{1}{1cm}{\textbf{Client}} & 
\multicolumn{3}{c|}{\textbf{Network-level (E1)}} &
\multicolumn{3}{c|}{\textbf{Colluding network-level (E3)}}&
\multicolumn{3}{c|}{\textbf{State-level (E4)}}\\\cline{2-10}
& \textbf{Websites (Main)} & \textbf{Websites (Any)} & \textbf{Circuits (All)} &
\textbf{Websites (Main)} & \textbf{Websites (Any)} & \textbf{Circuits (All)} & 
\textbf{Websites (Main)} & \textbf{Websites (Any)} & \textbf{Circuits (All)} \\\hline
\systemname & 3\% & 8\% & 2\% & 6\% & 13\% & 5\% & 27\% & 34\% & 25\%
\\\hline
Vanilla Tor & 37\% & 53\% & 40\% & 40\% & 56\% & 42\% & 82\% & 88\%
& 85\% \\\hline
\end{tabularx}
\caption{\edit{\systemname vs. vanilla Tor: An estimate of the threat faced from various 
attackers.\label{fig:astoria-security}}}
\end{table*}

\begin{figure*}[htb]
\centering
\begin{subfigure}[b]{0.45\textwidth}
\includegraphics[trim=0cm 0cm 0cm 2.5cm, clip=true,width=\textwidth]
{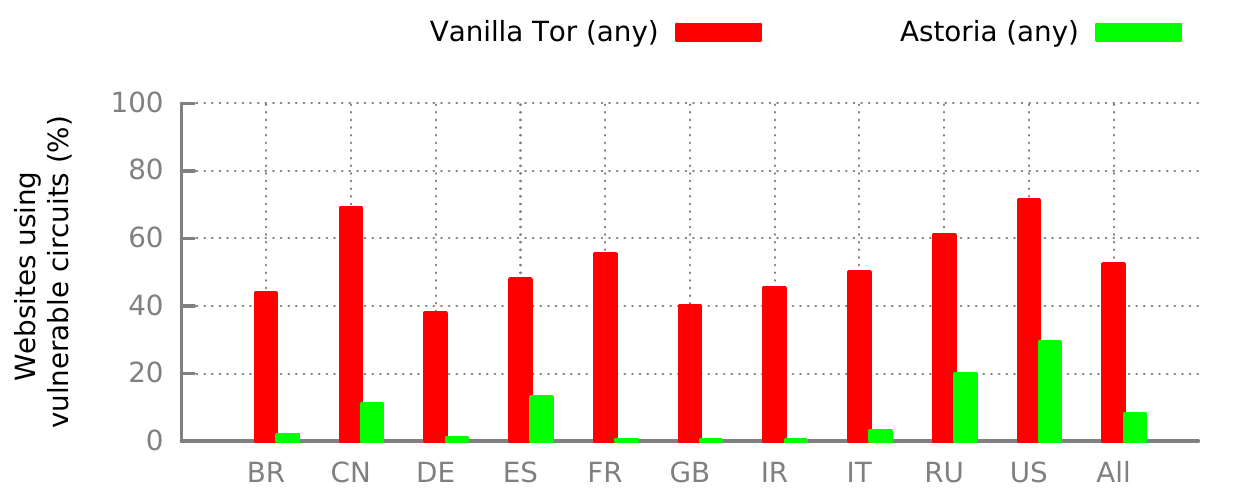}
\caption{Any request vs. Single AS adversaries [Experiment \textbf{E1}]}
\label{fig:any-as-level}
\end{subfigure}
\begin{subfigure}[b]{0.45\textwidth}
\includegraphics[trim=0cm 0cm 0cm 2.5cm, clip=true,width=\textwidth]
{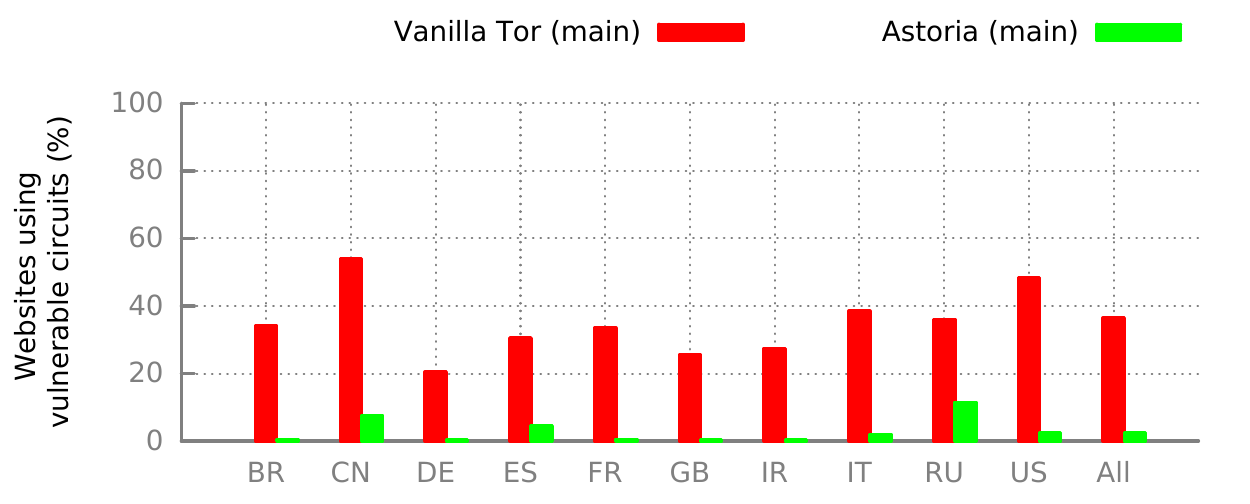}
\caption{Main request vs. Single AS adversaries [Experiment \textbf{E1}]}
\label{fig:main-as-level}
\end{subfigure}

\begin{subfigure}[b]{0.45\textwidth}
\includegraphics[trim=0cm 0cm 0cm 2.5cm, clip=true,width=\textwidth]
{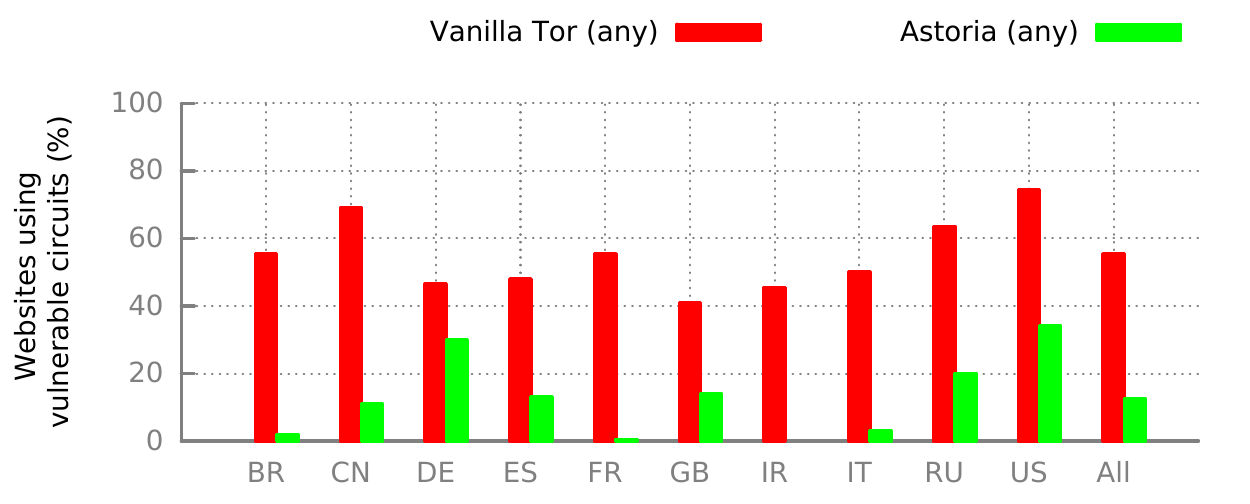}
\caption{Any request vs. Sibling AS adversaries [Experiment \textbf{E3}]}
\label{fig:any-colluding}
\end{subfigure}
\begin{subfigure}[b]{0.45\textwidth}
\includegraphics[trim=0cm 0cm 0cm 2.5cm, clip=true,width=\textwidth]
{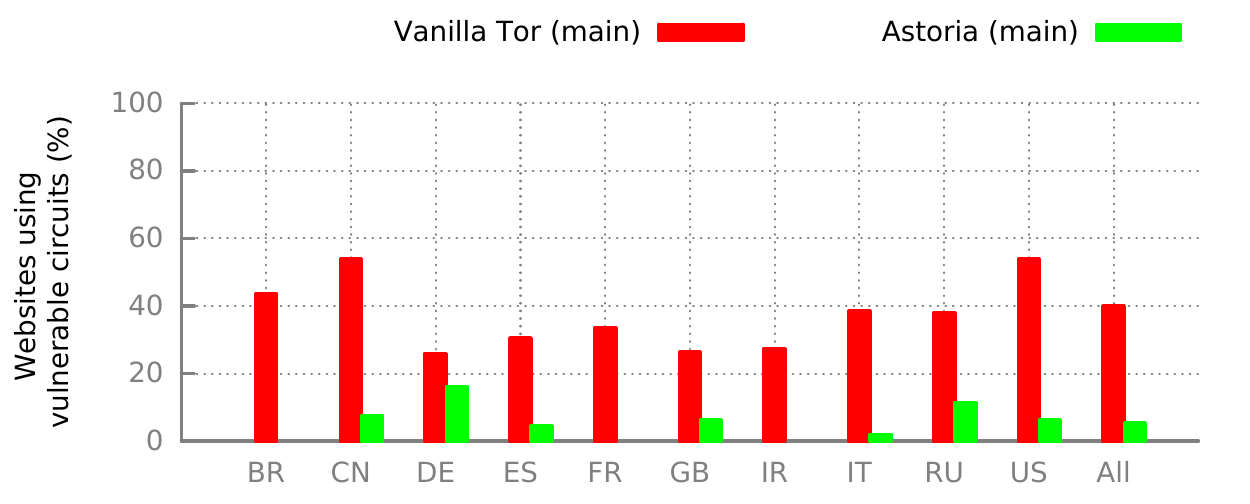}
\caption{Main request vs. Sibling AS adversaries [Experiment \textbf{E3}]}
\label{fig:main-colluding}
\end{subfigure}

\begin{subfigure}[b]{0.45\textwidth}
\includegraphics[trim=0cm 0cm 0cm 2.5cm, clip=true,width=\textwidth]
{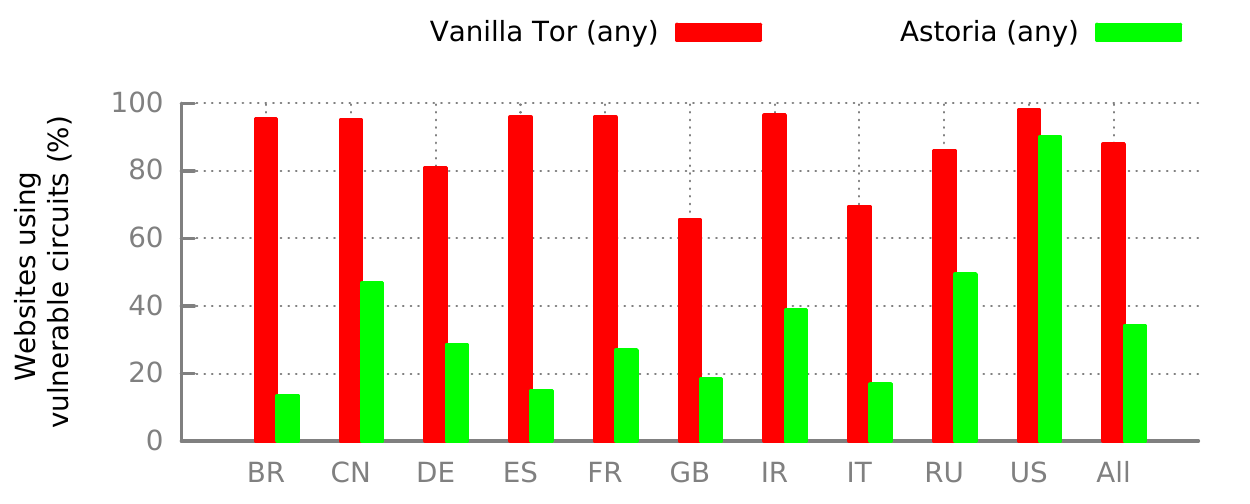}
\caption{Any request vs. State-level adversaries [Experiment \textbf{E4}]}
\label{fig:any-state}
\end{subfigure}
\begin{subfigure}[b]{0.45\textwidth}
\includegraphics[trim=0cm 0cm 0cm 2.5cm, clip=true,width=\textwidth]
{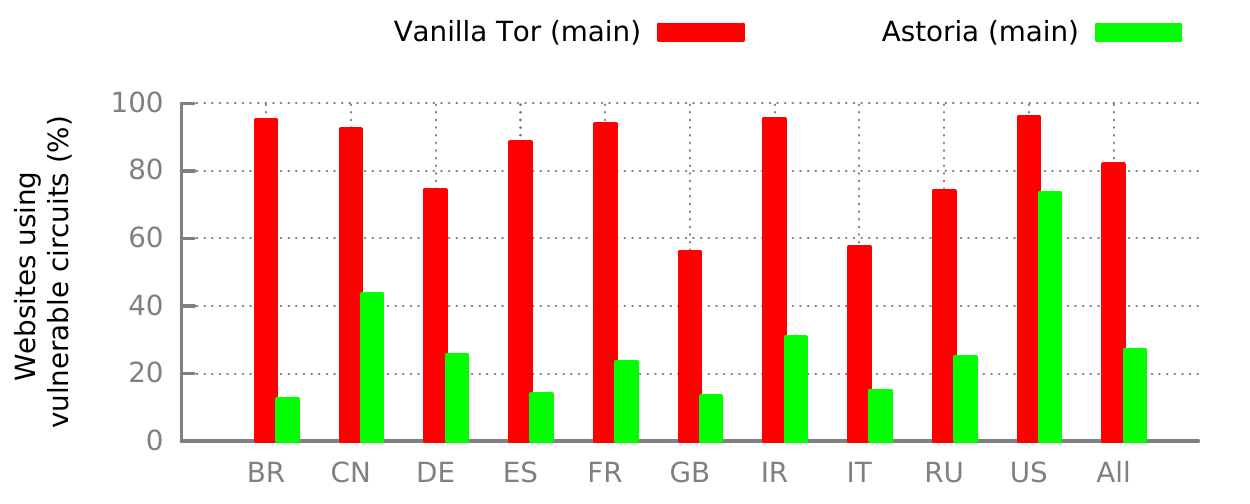}
\caption{Main request vs. State-level adversaries [Experiment \textbf{E4}]}
\label{fig:main-state}
\end{subfigure}

\caption{\edit{\systemname vs. vanilla Tor: Percentage of websites using vulnerable circuits for their main request or any request, against various adversaries.}}
\label{fig:ecdf-sim}
\end{figure*}

\myparab{E1: Measuring vulnerability to network-level attacks. }In this
experiment, we compare the security provided by the \systemname client with the
vanilla Tor client, against network-level adversaries. The threat from such
adversaries is significantly reduced from up to 40\% of all circuits being
vulnerable to 3\%, with the \systemname client. Figures 
\ref{fig:any-as-level} and \ref{fig:main-as-level} breaks down the results of this experiment by country. 
We see that \systemname completely removes the threat 
of network-level attackers on circuits carrying the main page request in 
clients from Brazil, France, and Iran, while bringing the risk down to under 
5\% in six other countries. 

%
%
%
%

\myparab{E3: Measuring the impact of sibling ASes. }We find that siblings have
little impact on the security provided by \systemname. Over all circuits
constructed by \systemname, the addition of colluding sibling ASes resulted in
less than a 3\% increase in number of vulnerable circuits, with the only
significant increase being in Germany (DE). This is illustrated in Figures
\ref{fig:any-colluding} and \ref{fig:main-colluding}. This large increase in number of vulnerable circuits 
indicates that if sibling ASes in Germany were to collude, \systemname (given 
the VPN client location and selected entry-guards) is often left with no safe 
entry- and exit- relay options for circuit construction. It is important to 
note that although there are a significant number of vulnerable circuits 
created by \systemname, these circuits are constructed using our linear program 
(Eq. \ref{eq:lp}) which minimizes the number of circuits visible to each
attacker.


\myparab{E4: Measuring the impact of state-level adversaries. }\systemname
performs reasonably well even against state-level adversaries by reducing the
fraction of potentially vulnerable circuits from 85\% (vanilla Tor) to 25\%, 
over all countries. The per country breakdown is illustrated in Figures 
\ref{fig:any-state} and \ref{fig:main-state}. The results show a steep decrease in the ratio of
vulnerable websites for all countries except the United States (US). This is due
to the large presence of American ASes on paths to and from our US VPN vantage
point and the entry-guards and any Tor exit-relay and our US destinations.


\myparab{Defending against active network-level attacks. }
\systemname focuses on adversaries who may lie on asymmetric network paths 
between the client and entry; and exit and destination, respectively. However, 
Sun~\etal~\cite{Sun-Arxiv15} highlight attacks based, not only on static path 
properties, but also dynamics of BGP (\eg hijacks, routing instability). Taking 
this sort of attack into account is challenging as it requires realtime access 
to interdomain routing data and intelligent analysis to identify incidents that
may impact the safety of the client's path. In the future, we plan to integrate
subscriptions to BGP hijack data sources (\eg Argus~\cite{argus}, or ongoing
efforts at building a real-time interception detector~\cite{caida-grant}) 
into \systemname to allow it to operate on dynamic BGP paths.

\subsection{Security against relay-level attackers}\label{subsec:relay-level}
In order to defend against relay-level attackers, \systemname inherits the
concept of entry-guards from the vanilla Tor client and also ensures that no 
two relays from the same family are placed on the same circuit. However, due 
to its AS-awareness, \systemname (and any AS-aware client that constructs 
circuits which are a function of the destination AS) currently is vulnerable to 
two relay-level attacks:
(1) it is possible for a middle-relay in an \systemname constructed circuit to
narrow down the set of possible (source, destination) AS pairs that are at
either end of the circuit (based on the selected entry- and exit-relays), and
(2) when \systemname is used from regions with no safe (entry, exit) relay
options, it is possible for a relay-level attacker to force \systemname to
create circuits that can be de-anonymized by it. Below, we discuss these 
attacks, their impact, and how to mitigate them. 

\myparab{Measuring the threat posed by middle-relays. }As seen in Table
\ref{fig:astoria-security}, in a majority of all cases,
\systemname is able to find a safe pair of entry- and exit-relays to use for
its circuits. As a result, an adversarial middle-relay working under the
assumption that \systemname always constructs safe circuits, will be able to
narrow down the set of possible source- and destination-ASes by simply
observing the entry- and exit-relays in the circuit. Below, using the results 
of experiment \textbf{E2} and statistical inference techniques, we show that 
the threat from such adversarial relays is negligible.

First, given our random sample of 100 source ASes for each country (and fixed
set of destinations) we infer the mean number of (source, destination) pairs
with greater than \textbf{50}\% safe entry- and exit-relay pair options for the
entire population of source ASes in each country (with the same fixed
destinations). Then, we find a lower-bound estimate on the expected number of
(source, destination) AS pairs that have each (entry, exit) pair as a safe
option -- i.e., a lower-bound on the number of (source, destination) pairs 
that can be linked to the circuit by a middle-relay in a single 
observation. Finally, we show that given the current
distribution of Tor relays, the probability of narrowing down this set of 
sources to a single (source, destination) pair is negligible.

\emph{Inferring the mean number of (source, destination) pairs with greater
than \textbf{50}\% safe options. }Recall that in experiment \textbf{E2}, 100
source ASes were selected at random from the set of all ASes in each country.
The experiment considers the destination ASes generated by the loading of 200
non-random destinations. Let the set of sampled source ASes be denoted by 
$\bar{X}$ and the set of destination ASes be denoted by $D$. From the results 
of the experiment, we extract the mean fraction of 
($\bar{x} \in \bar{X}$, $d \in D$) pairs which have more than \textbf{50}\% 
safe entry- and exit-relay options (denoted by $\mu_{\bar{X},D}$).
Let $X$ denote the set of all ASes within each country. Now, using the central
limit theorem and the sampling distribution of the sample means \cite{rice1995}, 
we infer the \textbf{99}\% confidence-interval for the mean fraction of
($x \in X$, $d \in D$) pairs which have more than \textbf{50}\% safe entry- and
exit-relay options (denoted by $\mu_{X,D}$).

\emph{Estimating a lower-bound on linkable sources. }We take an extremely
conservative approach to derive this lower-bound. First, we use the lower
value of $\mu_{X,D}$ from our \textbf{99}\% confidence interval. Further, we
assume that $\mu_{X,D}$ fraction of our ($x \in X$, $d \in D$) pairs have only
exactly \textbf{50}\% safe entry- and exit-relay options (although $\mu_{X,D}$
denotes the fraction of ($x \in X$, $d \in D$) pairs with greater than
\textbf{50}\% safe options). Finally, we assume that the remaining
$1-\mu_{X,D}$ fraction of ($x \in X$, $d \in D$) pairs have no safe options.
Given these assumptions, we can compute the lower-bound on the expected number
of $(x \in X, d \in D)$ pairs which have each (entry, exit) pair as a safe
option (denoted by $E[S_{en,ex}]$) as: $E[S_{en,ex}]$ =
$\frac{\text{Total safe circuits}}{\text{Total (entry, exit) pairs}}$ =
${.50 \times \mu_{X,D} \times |X| \times |D|}$.

$E[S_{en, ex}]$ is a lower-bound on the expected number of linkable source and
destination pairs for each observation of an entry- and exit-relay (under the
conservative assumption that an adversarial middle-relay knows the country in
which the client is located and the set of all possible destinations $D$ that
any client may connect to). 

\emph{Estimating the probability of complete de-anonymization. }Given that
$E[S_{en, ex}]$ is the number of $(x \in X, d \in D)$ pairs that are linkable
to a single observation of an (entry, exit) pair and assuming a constant rate
of reduction in linkable pairs (given by $\frac{E[S_{en, ex}]}{|X|\times|D|}$),
the number of circuits that need to be observed by the adversarial middle-relay
to narrow down the number of $(x \in X, d \in D)$ pairs to 1 -- i.e., to
completely de-anonymize the source and destination -- is
$n = \frac{-\log(|X|\times|D|)}{\log(E[S_{en,ex}])-\log(|X|\times|D|)}$
(since $(\frac{E[S_{en,ex}]}{|X|\times|D|})^n = \frac{1}{|X|\times|D|}$).

Since \systemname (1) constructs new circuits only if there are no existing
circuits that serve the same destination AS, and (2) selects middle-relays for
each new circuit according the the bandwidth distribution of relays, we obtain
the expected upper-bound of the probability of a middle-relay being able to 
observe $n$ circuits
between the same source and destination ASes (with different entry- and
exit-relays). Table \ref{tab:mean-infer} shows that this probability (denoted
by $P_n$) is negligible even for the Tor relay with the current highest
advertised bandwidth where the probability of selection as the middle-relay is
\textbf{.007}.

\begin{table}[!t]
\begin{tabularx}{.475\textwidth}{|l|l|l|l|X|X|l|X|}
\hline
\textbf{}&\textbf{$|X|$}&\textbf{$|D|$}&\textbf{$\mu_{\bar{X},D}$}&
{99\%CI $\mu_{X,D}$}&{E [$S$]}& \textbf{$n$} & \textbf{$P_{\lfloor n
\rfloor}$}\\\hline
BR      & 3,515  & 165   & .40  & (.39, .41)  & 114,797& 8.1  &5.7 $\times 
10^{-18}$\\\hline
CN      & 1,227  & 131   & .44  & (.43, .46)  & 35,216 & 7.8   &8.2 $\times
10^{-16}$       \\\hline
DE      & 2,022  & 190   & .33  & (.33, .34)  & 63,409 & 7.1   &8.2 $\times
10^{-16}$       \\\hline
ES      & 703   & 181   & .40  & (.39, .41)  & 25,295 & 7.2   &8.2 $\times
10^{-16}$       \\\hline
FR      & 1,251  & 187   & .32  & (.31, .33)  & 36,448 & 6.6   &1.1 $\times
10^{-13}$       \\\hline
GB      & 2,372  & 187   & .35  & (.34, .36)  & 76,473 & 7.3   &8.2 $\times
10^{-16}$       \\\hline
IR      & 470   & 133   & .39 & (.38, .40)  & 11,878 & 6.6   &1.1 $\times
10^{-13}$       \\\hline
IT      & 932   & 201   & .29 & (.28, .30)  & 26,800 & 6.2   &1.1 $\times
10^{-13}$       \\\hline
RU      & 5,868  & 178   & .27  & (.26, .28)  & 140,201& 6.9   &1.1 $\times
10^{-13}$       \\\hline
US      & 23,588 & 188   & .45  & (.44, .46)  & 977,768& 10.1  &2.8 $\times
10^{-22}$       \\
\hline
\end{tabularx}
\caption{Results from statistical analysis of the expected upper-bound of the 
threat posed by adversarial middle-relays on \systemname (using data obtained
from our simulation experiment (\textbf{E2}).}
\label{tab:mean-infer}
\end{table}

\myparab{Defending against attacks due to predictable relay-selection when
there are no safe options.}  
In certain client locations (\eg some ASes in China and Iran), there are no safe
entry- and exit-relay selections for some destinations, regardless of the guards
used by the client. In these cases, a relay-level adversary may place entry-and
exit-relays in ASes that provide a safe-path for \systemname clients attempting 
to connect to specific target destinations. This manipulates \systemname into 
using the adversarial (entry, exit) pair on all circuits connecting the client 
to the target destination -- allowing trivial de-anonymization of the user.

\systemname can defend against such attacks by selecting from safe (entry, exit) 
pairs only when a minimum threshold of available safe (entry, exit) pairs is
met. In cases where the threshold is not met, \systemname may discard the few
remaining safe pairs and choose entry- and exit-relays according to the
distribution produced by its linear program (Eq. \ref{eq:lp}), which minimizes
the amount of information gained by the network-level adversary. This however,
enables correlation attacks by selected network-level attackers. Since
it is not yet clear if network-level adversaries pose a larger threat than
relay-level adversaries. Therefore, determining this threshold is a non-trivial
open research problem.

\section{Discussion}\label{sec:discussion}
In this section, we compare the \systemname Tor client with the hypothetical
perfect Tor client and discuss how \systemname can be augmented and improved
with recent and ongoing developments from the network measurement community.

\subsection{Comparing \systemname and the perfect Tor client}
\label{sec:perfect} Here we point out some of the shortcomings of \systemname
when compared to the perfect Tor client. We find that many of these apply to
any AS-aware client. The perfect Tor client is able to simultaneously achieve 
three conflicting goals: 

\myparab{Defend against network-level attackers. }The perfect Tor client is 
able to prevent compromise from network-level attackers. In particular, the 
client constructs circuits that are safe from traffic correlation attacks. 

While such adversaries are largely ignored by the vanilla Tor client, 
\systemname successfully deals with them by utilizing efficient path-prediction 
tools to explicitly avoid relays that enable correlation attacks. However, 
\systemname does not currently deal with attacks from active network-level 
adversaries that are able to exploit BGP dynamics. In addition, 
\systemname is unable to exactly predict the paths that will be utilized to 
communicate with each Tor relay, and therefore only makes estimates (which are 
validated to be reasonably tight estimates).
 
\myparab{Defend against relay-level attackers. }Since the Tor network is
volunteer driven, it is critical for the perfect Tor client to be able to 
defend against passive and active attackers that are able to control a fraction 
of all relays within the network. This primarily involves (1) constructing circuits 
so that the probability of an adversarial pair of relays occupying the entry- 
and exit-hop of the circuit is low, and (2) ensuring that no single relay should 
be able to conclusively link the source and destination of the circuits it is 
on. 

While the vanilla Tor client is able to successfully mitigate threats from many
types of relay-level attacks, we find that this is challenging for AS-aware 
clients such as \systemname. First, while the concept of entry-guards mitigates 
many threats from relay-level attackers, it has a negative influence on the 
number of safe circuits that can be built by AS-aware clients. Second, AS-aware 
circuits inherently leak some information about the source and destination of 
the circuit. Our analysis in Section \ref{subsec:relay-level} shows that in the
average-case, \systemname circuits are safe from de-anonymization due to these
leaks.

\myparab{Maintain performance and load-balancing. }The perfect Tor client
must also perform load-balancing to ensure that no single set of relays in the
network are overloaded, while providing reasonable performance for all its
users.

In Section \ref{sec:evaluation} we demonstrated that \systemname performs
load-balancing in an identical manner to the vanilla Tor client and page-loads
are only slightly slower in most cases. There are two main reasons for
\systemname's increased page-load times: (1) Path prediction is expensive, and
(2) \systemname loses the ability to pre-emptively construct circuits. While (1)
is unavoidable, there are interesting future research questions regarding (2) --
\eg can smart caching and pre-emptive/predictive circuit construction for a set 
of popular/predicted destinations result in significant performance gains?

\subsection{Improving path-prediction accuracy}\label{subsec:paths}
Measuring the potential threat of
correlation attacks is made challenging by the fact that it requires measuring
both forward and reverse network paths between the client and entry, and exit
and destination, respectively. Thus, we opt to leverage an up-to-date map of the
Internet's topology, augmented with inferred business relationships between
networks and a model of routing policies to infer network paths. Modeling of
interdomain routing is a thorny issue and we take care to avoid well known
pitfalls including complex business relationships (e.g., ASes that act as a
customer in one geographic region, and a peer in others) and sibling ASes (ie.,
multiple ASes which correspond to a single organization). The issue of siblings
ASes is particularly relevant in our context, as multiple ASes controlled by a
single organization may share information to perform a correlation attack.
Despite all this, accurate path prediction remains an open challenge. In a
related study, we validate the accuracy of this approach and find that measured
paths follow this model 65-85\% of the time \cite{Anwar-TR15}. As a result, the 
numbers we observe should be taken as an estimate of the threat. 

We note that novel path measurement tools are on the
horizon (\eg Sibyl~\cite{sibyl}) that take into account richer vantage point
sets than prior work (\eg PlanetLab used by iPlane~\cite{iplane} \vs RIPE
Atlas~\cite{ripe-atlas} used by Sibyl). An interesting future direction is
determining how such measurement planes can be integrated into a Tor client (\eg
to operate in an offline mode or via a secured querying interface).

\section{Conclusions}\label{sec:conclusions}
We have leveraged highly-optimized algorithmic simulations of interdomain
routing on empirically-derived AS-level topologies to quantify the potential
for correlation attacks where an adversary can leverage asymmetric Internet
routing and collude with others within the same organization. Our results show
that a significant number of Tor circuits are vulnerable to AS- and state-level
attackers. 

To mitigate the threat from such attackers, we developed \systemname---an 
AS-aware Tor client. Beyond providing a high-level of security against these 
attacks, \systemname also has performance that is within a reasonable distance 
from the current Tor client. Also, unlike other AS-aware Tor clients,
\systemname also considers how circuits should be built in the worst case, 
i.e., when there are no safe relays available to the client. Further,
\systemname is a good network citizen and is designed to ensure that the all 
circuits created by it are load-balanced across the volunteer-driven Tor 
network. 

Our work highlights the importance of applying current models and data from
network measurements to inform relay selection so as to protect against timing
attacks. \systemname also opens multiple avenues for future work such as
integrating real-time hijack and interception detection systems (to fully
counter RAPTOR ~\cite{Sun-Arxiv15} attacks) and understanding how new 
measurement services can be leveraged by a Tor client without defeating 
anonymity.

\myparab{Source code: }The source code of the \systemname client is available under the CRAPL
\footnote{\url{http://matt.might.net/articles/crapl/}} 
license at \url{http://nrg.cs.stonybrook.edu/astoria-as-aware-relay-selection-for-tor/}.

\section*{Acknowledgments}
We would like to thank Ruwaifa Anwar, Haseeb Niaz, and Abbas Razaghpanah for
their help with integrating sibling detection algorithms into our measurement 
toolkit. 
 
This material is based upon work supported by the National Science
Foundation under Grant No. CNS-1350720, a Google Faculty Research Award, ISF
grant 420/12, Israel Ministry of Science Grant 3-9772, Marie Curie Career
Integration Grant, Israeli Center for Research Excellence in Algorithms
(I-CORE), and an Open Technology Fund Emerging Technology Fellowship. 
Any opinions, findings, and conclusions or recommendations expressed in
this material are those of the author(s) and do not necessarily reflect the
views of the National Science Foundation, Google, the Israel Ministry of
Science, or the Open Technology Fund.

\bibliographystyle{IEEEtranS}
\bibliography{IEEEabrv,intercept}

\end{document}